\pdfoutput=1

\documentclass[a4paper,11pt]{article}
\pdfoutput=1 

\usepackage{jcappub} 

\usepackage[T1]{fontenc} 
\usepackage[utf8x]{inputenc}
\usepackage[english]{babel}
\usepackage{subfigure}
\usepackage[nottoc]{tocbibind}
\usepackage{booktabs}

\usepackage{xspace}
\usepackage{tabularx}
\usepackage{multirow}
\usepackage{enumitem}
\usepackage{longtable}
\usepackage{xcolor}
\usepackage{multirow}
\usepackage{soul}
\usepackage{xcolor}
\usepackage{appendix}
\usepackage[normalem]{ulem}
\usepackage{wrapfig}
\usepackage{lscape}
\usepackage{rotating}
\usepackage{epstopdf}
\usepackage{cancel}
\usepackage{listings}
\usepackage{hyperref}
\hypersetup{pdfpagemode={UseOutlines},bookmarksopen,pdfstartview={FitH},colorlinks,linkcolor={blue},citecolor={blue},urlcolor={blue}}

\newcommand{\vbc}{v_{\text{bc}}}
\newcommand{\fFDM}{f_{\rm FDM}}

\definecolor{darkblue}{RGB}{14,0,185}
\definecolor{darkred}{RGB}{175,0,0}

\title{
Dark ages, a window on the dark sector. \\
Hunting for ultra-light axions
}

\author[a,b]{Eleonora Vanzan,}
\emailAdd{eleonora.vanzan@phd.unipd.it}

\author[a,b,c]{Alvise Raccanelli,}
\emailAdd{alvise.raccanelli.1@unipd.it}

\author[a,b,c]{Nicola Bartolo}
\emailAdd{nicola.bartolo@pd.infn.it}

\affiliation[a]{Dipartimento di Fisica e Astronomia ``Galileo Galilei'', Universit\` a degli Studi di Padova, via Marzolo 8, I-35131 Padova, Italy}
\affiliation[b]{INFN Sezione di Padova, via Marzolo 8, I-35131 Padova, Italy}
\affiliation[c]{INAF, Osservatorio Astronomico di Padova, vicolo dell’Osservatorio 5, I-35122 Padova, Italy}

\hypersetup{pdfauthor={Eleonora Vanzan}}

\begin{abstract}
{
Measurements of 21cm intensity mapping (IM) during the dark ages can potentially provide us with an unprecedented window on high redshifts and small scales.
One of the main advantages this can bring involves the possibility to probe the nature of dark matter.
Tests of dark matter models with the large-scale structure of the Universe are limited by non-linearities and astrophysical effects, which are not present for IM measurements during the dark ages.
In this paper we focus on constraining the model in which dark matter is comprised, totally or in part, by ultra-light axion-like particles around the $10^{-18}-10^{-22}$ eV mass scale.
For this model, the angular power spectrum of 21cm brightness temperature fluctuations will exhibit a small-scale suppression.
However, this effect is intertwined with the imprint of baryon-dark matter relative velocity at recombination, causing at the same time an enhancement at large-scales, which is affected by the mass and abundance of axion dark matter.
In this work we forecast how future radio arrays will be able to constrain ultra-light axion mass through both these effects on the angular power spectrum.
}
\end{abstract}

\begin{document}

\maketitle

\section{Introduction}
\label{sec:intro}
Among the many unanswered questions in cosmology, the nature of dark matter, which makes up roughly 25\% of the present energy budget of the Universe, is one of the big elephants in the room.
The standard cold dark matter (CDM) scenario has been tested on large scales and provides a good fit to the data~\cite{Bertone2004}. However, below galactic scales, there are some inconsistencies with observations, such as the ``missing satellite'' problem and the ``cusp-core problem''~\cite{Bullock2017}. These puzzles seem to hint to a lack of power with respect to the standard $\Lambda$CDM picture.

A promising alternative candidate is in the form of ultra-light axion dark matter (ULAs) or fuzzy dark matter (FDM), which consists in dark matter being made of ultra-light scalar fields with masses in the range $10^{-27} \sim 10^{-10} \text{ eV}$~\cite{Hu2000, Marsh2015review, Hui2016}. This scenario provides the same large-scale predictions as CDM, but differs on small scales due to the wave-like nature of the fields: such light particles have a large, macroscopic de Broglie wavelength which suppresses small-scale structure. The new crucial ingredient is quantum pressure: its effects enter into play on scales comparable to the axion Jeans scale $k_{J,a} \sim a\sqrt{Hm_a}$, and this induces a suppression on the matter power spectrum with respect to the CDM case. The lighter the field, the lowest the wavenumber where it suppresses power.

A large number of ultra-light axion fields, with masses down to the Hubble scale, $10^{-33} \text{ eV}$, arise e.g.,~in axiverse scenarios~\cite{Arvanitaki2009, Kamionkowski2014, Visinelli2018}.
A mass range similar to the one investigated in this work has been analyzed in~\cite{ Sarkar2022} and~\cite{Hotinli2022} by looking at cosmic dawn measurements.
Heavier masses in the FDM window have been studied through black hole superradiance~\cite{Brito2015}, while Ly$\alpha$ constraints have been shrinking the allowed window from above~\cite{Viel2013, Kobayashi2017, Rogers2020}. However, if one allows FDM to constitute just a fraction $\fFDM$ of the totality of dark matter, the allowed mass range still spans many orders of magnitude, see~\cite{Flitter2022}. It is possible for FDM to exist in large portions in the region $10^{-25}$ eV -- $10^{-23}$ eV, while for lighter masses current constraints suggest $\fFDM \lesssim 10^{-2}$, and for heavier masses $\fFDM \lesssim 10^{-1}$. In the mass range $10^{-32} \text{ eV} \sim 10^{-25.5} \text{ eV}$, CMB and LSS data impose that the ULAs relic density obeys $\Omega_a/\Omega_d \leq 0.05$ and $\Omega_a h^2 \leq 0.006$~\cite{Hlozek2014,Hlozek2017}. In the mass window $10^{-25} \text{ eV} \sim 10^{-23} \text{ eV}$, future experiments such as HERA will be sensitive to $\fFDM$ of order 0.01~\cite{Flitter2022}. Even if ultra-light axions were to be only a fraction of the dark matter, it would still be possible for the whole dark matter to be comprised of axions spanning a much wider mass range, in an axiverse-like scenario with an extended mass function~\cite{Acharya2010, Acharya2010a, Mehta2021}.

A novel way to probe the nature of dark matter, which is maturing in the last years, is given by measurements of 21cm line intensity mapping (see e.g.~\cite{Kovetz2017} for a recent review). This observable has the potential to open an unprecedented window on high redshifts $30 \lesssim z \lesssim 200$, shedding light on the dark ages of the Universe, i.e.,~the epoch between recombination, when neutral hydrogen is formed, and CMB photons can freely stream, and the formation of the first stars at $z \lesssim 30$.

In order to test the fuzzy dark matter scenario, one would like to investigate the matter power spectrum at small scales. For 21cm intensity mapping, this would mean observing its angular power spectrum up to $\ell \sim \mathcal{O}(10^6,10^7)$, which is still very futuristic. However, there are second order perturbative effects that cause low multipoles enhancements, induced by the relative velocity between baryons and dark matter $\vbc$ at recombination~\cite{Tseliakhovich2010}: depending on the mass of the axion, these effects could leave a detectable imprint.

This relative velocity delays the growth of the first structures and therefore suppresses the matter power spectrum on scales around $\sim 10-10^3 \text{ Mpc}^{-1}$. This gives rise to additional effects on the 21cm angular power spectrum~\cite{Ali-Haimoud2013}. In particular, there is a long-short mode coupling that arises from having different patches in the early Universe with different values of the background relative velocity: the power spectrum gets modulated on the scales over which $\vbc$ varies $\sim 0.005 - 1$ Mpc$^{-1}$, and this leads to an enhancement of the power on the largest angular scales.

It has been pointed out in~\cite{Marsh2015} that an interesting interplay arises when the mass of the axion is such that the relative velocity suppression scales are comparable to the scales at which the quantum pressure effects kick in. Depending on the mass, ULAs may wash out power in the matter spectrum before or after the relative velocity suppresses the growth of structures: if axions are heavy enough, then the damping will happen at sufficiently high wavenumbers and the $\vbc$ effect will be present, consequently also the second order enhancement effect will be visible at large angular scales.
If, on the contrary, axions are too light, then they will wash out the relative velocity features at small scales and there will be no corresponding second order effect on large scales. For axion masses around $10^{-19} \text{ eV}$, ULAs and $\vbc$ effects come into play around the same scales.

For the first time, this work provides a quantitative forecast of the ability of future 21cm experiments in the dark ages to put constraints on the mass of ULAs, taking advantage of the interplay between axion effects and relative velocity effects on the matter power spectrum. While both $\vbc$ and ULAs damping are appreciable at very high multipoles, of order $\sim 10^6$, searching for second order effects at low $\ell$s is a much more realistic avenue and it drastically improves the constraining power, even for ground based experiments such as an hypothetical advanced version of the SKAO.

The structure of the paper is as follows. Sections~\ref{sec:21cm}-\ref{sec:ulas} give a very brief review of 21cm physics, of the dark matter-baryon relative velocity, of the second order imprints of $\vbc$ on the 21cm angular power spectrum, and of ULAs. Section~\ref{sec:analysis} outlines the methodology for the analysis. Finally, Section~\ref{sec:results} forecasts the ability of planned surveys and of more futuristic instruments to discriminate between CDM and ULAs, and discusses the relevant requirements for future surveys and the possible degeneracies among parameters.
The reference cosmology is taken from Planck2018~\cite{Planck2018}, and it will be: $\{ \omega_b=0.02238, \  \omega_{cdm}=0.1201, \  h=0.6781, \  n_s=0.9660, \  \ln 10^{10} A_s=3.045 \}$.

\section{21cm physics}
\label{sec:21cm}
This section very briefly reviews the physics of 21cm line intensity mapping; for some seminal papers and recent reviews, see e.g.,~\cite{Loeb2003, Zaldarriaga2003, Furlanetto2006, Lewis2007, Pritchard2008, Furlanetto2009, Pritchard2010, Pritchard2011, Kovetz2017, Furlanetto2019b, Bernal2022}.

The dark ages (generally defined as in the epoch $30 \lesssim z \lesssim 200$) are particularly interesting from a cosmological point of view, because they are not yet affected by astrophysical complications following the formation of the first luminous objects~\cite{Madau1996}.
During the dark ages, the Universe is filled with a neutral hydrogen gas: occasionally, the background CMB photons scatter on the hydrogen atoms and excite the hyperfine state of the electrons. A 21 cm wavelength photon is emitted as a result of the hyperfine transition of the hydrogen atom from the triplet state, where the spins of the electron and the proton are aligned, to the singlet state, where the spins are anti-aligned. Once the hydrogen gas has cooled sufficiently and its temperature is decoupled from the CMB temperature, this 21cm signal can be observed in absorption or emission with respect to the background of CMB photons. The energy difference between the two states, $E_{10} \approx 0.068$ K, corresponds to an emitted photon frequency $\nu_{21} \approx 1420$ MHz, which gets redshifted due to the expansion of the Universe, so that observing a given frequency singles out a unique redshift slice. This allows to perform a tomographic analysis, as if one had many CMB-like screens, one for each redshift slice.

Unfortunately, Earth's ionosphere is opaque to the redshifted 21cm signal emitted at $z \gtrsim 30$. Probing this era will require an observatory in space or on the far side of the Moon, where also radio frequency interference is minimized~\cite{Silk2020}. Being able to observe the 21cm signal in the dark ages would provide crucial advantages: on the one hand, fluctuations are not affected by Silk damping and remain undamped down to the baryon Jeans scale ($k \sim 300 \text{ Mpc}^{-1}$, to be compared to the photon diffusion scale $k \sim 0.2 \text{ Mpc}^{-1}$); on the other hand, line intensity mapping will enable us to observe the Universe in tomography. These two aspects imply that one could in principle probe a wide range of modes, extending well beyond those accessible to CMB experiments.

The spin temperature $T_s$ is defined from the ratio of abundances of neutral hydrogen in the triplet state $n_1$ and in the siglet state $n_0$
\begin{equation}
    \frac{n_1}{n_0} \equiv 3 e^{-E_{10} / T_s} \approx 3\left(1-\frac{E_{10}}{T_s}\right) \, .
\end{equation}
The spin temperature is determined by a balance: collisional transitions tend to set $T_s \xrightarrow{} T_{\rm gas}$, while radiative transitions mediated by CMB photons tend to set $T_s \xrightarrow{} T_{\rm CMB}$. During the dark ages, around $z \sim 100$, collisions efficiently couple the spin temperature to the gas temperature so that $T_s \approx T_{\rm gas}$, and the hydrogen emits or absorbs photons from the CMB when the local spin temperature is respectively higher or lower than the CMB one.

The local brightness temperature contrast with respect to CMB is
\begin{equation}
    T_{21}^{\rm loc} = (T_s-T_{\rm CMB}) (1-e^{-\tau}) \, ,
\end{equation}
where $\tau$ is the Sobolev optical depth
\begin{equation}
    \tau = \frac{3}{32\pi} \frac{E_{10}}{T_s} x_{HI} n_H \lambda_{21}^3 \frac{A_{10}}{H(z) +(1+z) \partial_r v} \, .
\end{equation}
with $x_{HI}$ the fraction of neutral hydrogen, $A_{10} \approx 2.85 \cdot 10^{-15}$ s$^{-1}$ the spontaneous decay rate, and $\partial_r v$ the line-of-sight gradient of the component of the peculiar velocity along the line of sight\footnote{This term $\partial_r v$ can be read as a perturbation of the Hubble expansion rate at the absorber's location. Although similar, it is not a redshift-distortion term. See the comment in~\cite{Ali-Haimoud2013}.}. Then the redshifted signal is
\begin{equation}
    T_{21}^{\rm obs} = \frac{T_{21}^{\rm loc}}{1+z} = \tau \frac{T_s-T_{\rm CMB}}{1+z} \, .
\end{equation}

Now the brightness temperature fluctuations are to be linked with the perturbations in the local hydrogen density and gas temperature.
The dependence of 21cm brightness temperature can be parametrized as:
\begin{equation}
    T_{21} = \Bar{T}_{21} \left( 1+\delta_v+\delta_v^2 \right) +\left( \mathcal{T}_b\delta_b +\mathcal{T}_T\delta_{T_{\rm gas}} \right)(1+\delta_v) +\mathcal{T}_{bb}\delta_b^2 +\mathcal{T}_{bT}\delta_b\delta_{T_{\rm gas}} +\mathcal{T}_{TT}\delta_{T_{\rm gas}}^2 \, ,
\end{equation}
with $\delta_v = -(1+z)\partial_rv_r/H(z)$ and $\delta_b=\delta n_b/\Bar{n}_b$, and neglecting $\delta_{x_e}$. Recall that $\delta_b=\delta_H$ up to negligible corrections. The mean brightness temperature $\Bar{T}_{21}$ is defined by setting all perturbations to zero, and all the coefficients $\mathcal{T}$ are all functions of redshift only.\\
It follows that, up to second order in fluctuations~\cite{Ali-Haimoud2013, Munoz2015},
\begin{equation}
\label{eq:T21fluctuations}
    \delta T_{21} = \mathcal{T}_H \delta_H +\mathcal{T}_T \delta_{T_{\rm gas}} -\Bar{T}_{21} \delta_v +\mathcal{T}_{HH} \left(\delta_H\right)^2 +\mathcal{T}_{TT} \left(\delta_{T_{\rm gas}}\right)^2 +\mathcal{T}_{HT} \delta_H\delta_{T_{\rm gas}} \, .
\end{equation}

\subsection{Dark matter-baryon relative velocity}

\begin{figure}[h!]
    \centering
    \includegraphics[width=.6\textwidth]{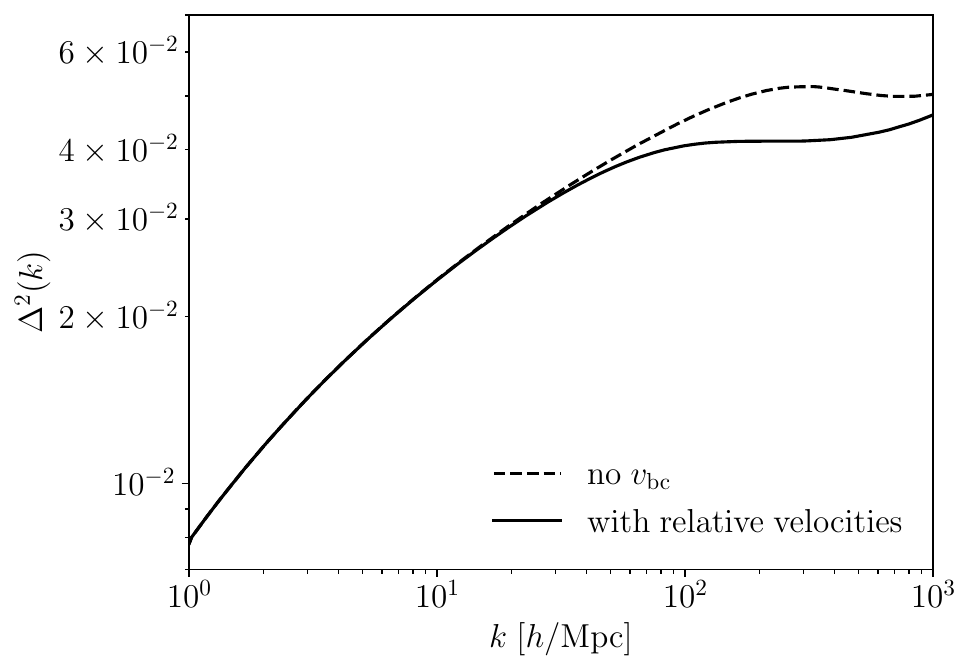}
    \caption{Relative velocity effect on the matter power spectrum, reproducing~\cite{Tseliakhovich2010}.}
    \label{fig:Pk_vbcEffect_z30}
\end{figure}

At the time of recombination, there is a highly supersonic relative velocity $\vbc$ between baryons and cold dark matter, of order 30 km/s~\cite{Tseliakhovich2010}. Indeed, baryons suffer Thomson scattering as long as they are tied to the photons, while dark matter travels along geodesics and starts to form gravitational potential wells. When baryons are no longer coupled to the photons, their sound speed drops, but they do not immediately fall into the dark matter potential wells: instead, the relative motion allows baryons to advect out of the wells and significantly suppresses the growth of structures on scales
\begin{equation}
    k_{\vbc} \equiv \left. \frac{aH}{\sqrt{\langle\vbc^2\rangle}} \right|_{\text{decoupling}} \sim 30-40 \text{ Mpc}^{-1} \, .
\end{equation}
In particular, the power is most strongly suppressed around the Jeans scale $k_J=aH/c_s \sim 200$ Mpc$^{-1}$, with a difference of $\sim 15\%$ with respect to the standard power spectrum, as shown in Figure~\ref{fig:Pk_vbcEffect_z30}.

In order to study the effect of relative velocities, the key observation is, there is a separation of scales. The largest scales, which were outside the sound horizon at decoupling, are left unaffected. The velocity field has a coherence length of about several comoving Mpcs. The relevant scales for the collapse and formation of structure correspond to a few baryonic Jeans lengths, that is $\sim 10$ comoving kpc. Thus the Universe can be pictured as being made of many different patches, of size comparable to the coherence length of the relative velocity: inside each of these patches, the relative velocity can take a different background value $\Vec{v}_{\rm bc}^{\text{(bg)}}$. The quantities of interest need to be averaged over all these regions.\\
If $k_{\rm coh}$ is the scale associated to the coherence length of $\vbc$, then this separation of scales
\begin{equation}
    k_l \lesssim k_{\rm coh} \ll k_{\vbc} \lesssim k_s
\end{equation}
allows to rewrite the fluid equations in the framework of moving background perturbation theory (MBPT)~\cite{Tseliakhovich2010}:
\begin{align}
    & \frac{\partial \delta_c}{\partial t} +\frac{1}{a}\Vec{v}_c\cdot\nabla \delta_c = -\frac{1}{a}\left(1+\delta_c\right) \nabla\cdot\Vec{v}_c \, , \label{continuity for cdm} \\
    & \frac{\partial \Vec{v}_c}{\partial t} +\frac{1}{a}\left(\Vec{v}_c\cdot\nabla\right)\Vec{v}_c = -\frac{1}{a}\nabla\Phi -H\Vec{v}_c \, , \label{Euler for cdm} \\
    & \frac{\partial \delta_b}{\partial t} +\frac{1}{a}\Vec{v}_b\cdot\nabla \delta_b = -\frac{1}{a}\left(1+\delta_b\right) \nabla\cdot\Vec{v}_b \, , \label{continuity for baryons} \\
    & \frac{\partial \Vec{v}_b}{\partial t} +\frac{1}{a}\left(\Vec{v}_b\cdot\nabla\right)\Vec{v}_b = -\frac{1}{a}\nabla\Phi -H\Vec{v}_b -\frac{c_s^2}{a}\nabla\delta_b \, , \label{Euler for baryons} \\
    & \nabla^2\Phi = 4\pi G a^2 \Bar{\rho}_m \delta_m \, .
\label{Poisson}
\end{align}
In the absence of density perturbations, but in the presence of a bulk velocity, there exists an exact solution:
\begin{equation}
\begin{split}
    \label{moving background solution}
    & \Vec{v}_c(\Vec{x},t) = \Vec{v}_c^{\text{(bg)}}(t) \, , \\
    & \Vec{v}_b(\Vec{x},t) = \Vec{v}_b^{\text{(bg)}}(t) \, , \\
    & \Phi = \delta_c = \delta_b = 0 \, ,
\end{split}
\end{equation}
where the background velocities decrease as $1/a(t)$. One has to perturb around~\eqref{moving background solution}
\begin{equation}
    \Vec{v}_b(\Vec{x},t) = \Vec{v}_b^{\text{(bg)}}(t) +\Vec{u}_b(\Vec{x},t) \, ,
\end{equation}
and the new perturbation variables are $\left\{ \delta_c, \Vec{u}_c, \delta_b, \Vec{u}_b, \Phi \right\}$.

Introducing the velocity divergence $\theta=\frac{1}{a}\nabla\cdot\Vec{v}$, and working in the bulk baryon frame, setting $\Vec{v}_b^{\text{(bg)}}=0$ and $\Vec{v}_c^{\text{(bg)}} = -\Vec{v}_{\text{bc}}^{\text{(bg)}}(t)$, the equations are
\begin{subequations}
\label{MBPT set of equations}
\begin{align}
    & \frac{\partial \delta_c}{\partial t} = \frac{i}{a}\Vec{v}_{\text{bc}}^{\text{(bg)}}\cdot \Vec{k}\delta_c -\theta_c \, , \\
    & \frac{\partial \theta_c}{\partial t} = \frac{i}{a}\Vec{v}_{\text{bc}}^{\text{(bg)}}\cdot \Vec{k}\theta_c -\frac{3H^2}{2}\left(\Omega_c\delta_c+\Omega_b\delta_b\right) -2H\theta_c \, , \\
    & \frac{\partial \delta_b}{\partial t} = -\theta_b \, , \\
    & \frac{\partial \theta_b}{\partial t} = -\frac{3H^2}{2}\left(\Omega_c\delta_c+\Omega_b\delta_b\right) -2H\theta_b +\frac{c_s^2k^2}{a^2}\delta_b \, .
\end{align}
\end{subequations}

\subsection{Second order effects on the 21cm power spectrum}

The fact that baryons and CDM have a supersonic relative velocity after recombination modifies the theoretical 21cm power spectrum \emph{on all scales}~\cite{Ali-Haimoud2013}.
In particular, the effect that is most relevant for the present work is a large-scale enhancement at $k \sim 0.005-1 \text{ Mpc}^{-1}$ of 21cm fluctuations. Two main ingredients enter the game. First, the relation between 21cm intensity and underlying baryonic fluctuations is non-linear, schematically $\delta T_{21} \approx \alpha \delta + \beta \delta^2$, with $\alpha$ and $\beta$ of comparable magnitude. Then on large scales $\delta T_{21,l} \approx \alpha \delta_l + \beta (\delta^2)_l$, and normally this second term would be negligible, for Gaussian initial conditions and while perturbations are still in the linear regime. As a second ingredient, however, the relative velocity $\vbc$ leads to a large-scale modulation of the amplitude of small-scale fluctuations, consequently $(\delta^2)_l \sim \delta_s^2$. Small-scale fluctuations are much larger than large-scale ones, $\delta_l \ll \delta_s \ll 1$, and the usually neglected quadratic term becomes comparable to the linear one $\delta_s^2 \sim \delta_l$, which leads to an enhancement of the large-scale 21cm power spectrum. Summarizing, the large-scale enhancement is a non-linear effect, which requires both the non-linear dependence and the fact that the linear and quadratic terms are of comparable magnitude.

Crucially, this allows to extract information on the behaviour of the small-scale power spectrum by looking at large scales.

For 21cm LIM, the MBPT must be complemented with the evolution equations for the gas temperature perturbations $\delta_{T_{\rm gas}}$ and the ionization fraction perturbations $\delta_{x_e}$. Following~\cite{Ali-Haimoud2013}, one can isolate in equation~\eqref{eq:T21fluctuations} a monopole source
\begin{equation}
    \delta_s \equiv \frac{\mathcal{T}_H\delta_H^{(1)} +\mathcal{T}_T\delta_{T_{\rm gas}}^{(1)}}{\Bar{T}_{21}} \, ,
\end{equation}
and a total quadratic term
\begin{equation}
    \delta T_{21}^{(2)} \equiv \mathcal{T}_{HH} \left(\delta_H^{(1)}\right)^2 +\mathcal{T}_{TT} \left(\delta_{T_{\rm gas}}^{(1)}\right)^2 +\mathcal{T}_{HT} \delta_H^{(1)} \delta_{T_{\rm gas}}^{(1)} +\mathcal{T}_T \delta_{T_{\rm gas}}^{(2)} \, .
\end{equation}
Then
\begin{equation}
    \delta T_{21}^{\rm obs} = \Bar{T}_{21} \left( \delta_s-\delta_v \right) +\delta T_{21}^{(2)} \, .
\end{equation}
The term $\delta T_{21}^{(2)}$ contains the total contribution of quadratic quantities. The procedure to compute it is described in detail in~\cite{Ali-Haimoud2013}, in particular Section III deals with the $\vbc$-induced, long-wavelength fluctuations of small-scale quantities. The small-scale perturbations are functions of the local value of the relative velocity, therefore their long-wavelength fluctuation is computed by taking a spatial smoothing over an intermediate scale $k_{\rm coh} \ll k_{\rm smooth} \ll k_{\vbc}$, which practically corresponds to averaging over the distribution of relative velocities.

The angular power spectrum of brightness temperature fluctuations is
\begin{equation}
\begin{split}
    &C_{\ell}(z) = e^{-2\tau_{\rm reion}} 4\pi \int\frac{d^3\Vec{k}}{(2\pi)^3} \left[  P_0(k,z) \alpha_{\ell}(k,z)^2 \right. \\
    &\quad \left. +2P_{0v}(k,\mu,z) \alpha_{\ell}(k,z) \beta_{\ell}(k,z) +P_v(k,\mu,z) \beta_{\ell}(k,z)^2 \right] \, ,
\end{split}
\end{equation}
where $P_0$ is the power spectrum of those terms that do not depend on the line of sight $\Bar{T}_{21}\delta_s+\delta T_{21}^{(2)}$, $P_v$ is the power spectrum of $\theta_b/H$, $P_{0v}$ is the cross-spectrum, and
\begin{align}
    & \alpha_{\ell}(k,z) \equiv \int dr j_{\ell}(kr) W(r,z) \, , \\
    & \beta_{\ell}(k,z) \equiv \int dr j^{\prime\prime}_{\ell}(kr) W(r,z) \, .
\end{align}

\begin{figure}[h!]
    \centering
    \includegraphics[width=0.45\textwidth]{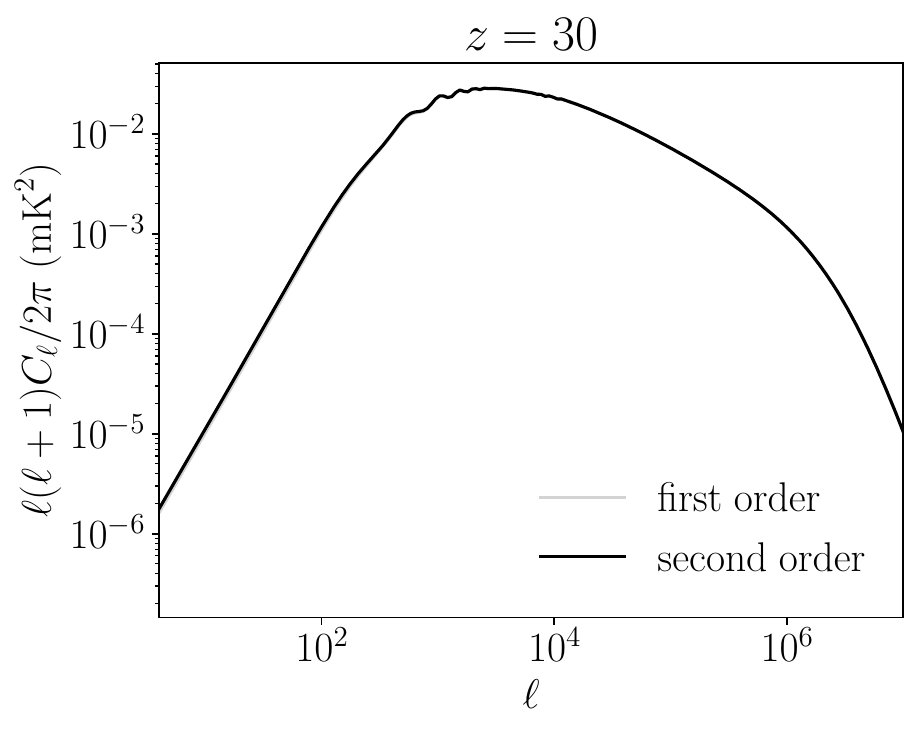}
    \includegraphics[width=0.45\textwidth]{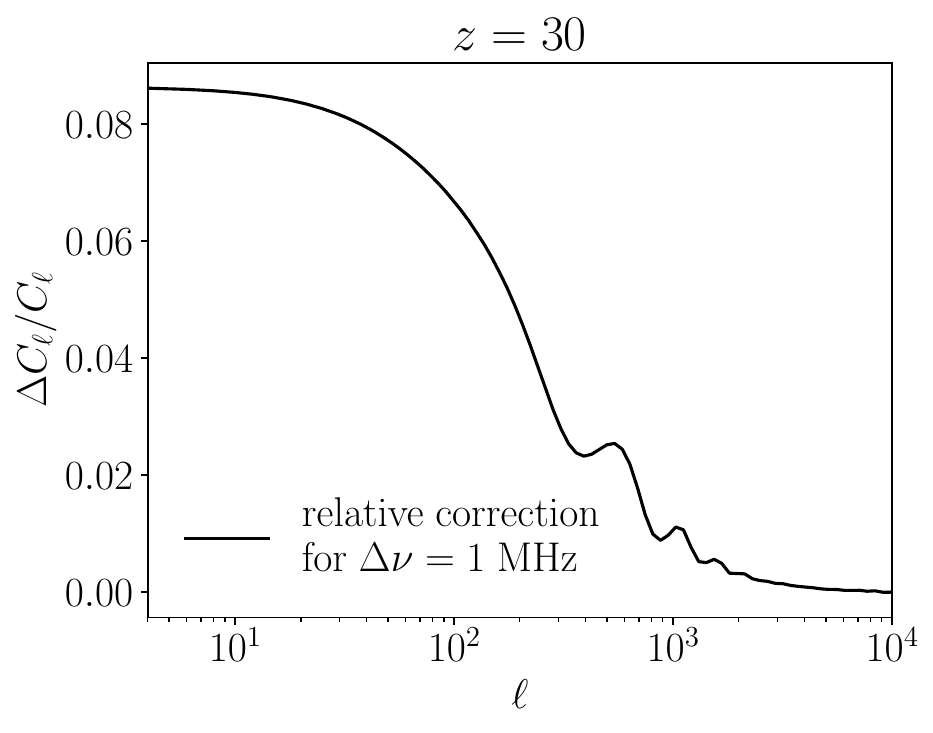}
    \caption{Second order effects and relative correction, reproducing Figure 15 in~\cite{Ali-Haimoud2013}.}
    \label{fig:Cls_CDM_1st&2nd}
\end{figure}

The second order enhancement on the 21cm angular power spectrum at low multipoles is of order few percent, as shown in Figure~\ref{fig:Cls_CDM_1st&2nd}.

\section{Ultra-light axions}
\label{sec:ulas}

Ultra-light axions, with masses around $10^{-20}$ eV, are appealing dark matter candidates because they alleviate some of the cold dark matter puzzles on scales $\lesssim 10$ kpc, such as the ``too big to fail'' problem or the cusp-core problem~\cite{Hu2000,Hui2016}, and they provide a new perspective on the $S_8$ tension~\cite{Rogers2020, Amon2022,Preston2023}.
With respect to standard cold dark matter, the main feature of ultra-light axions is their astrophysically large Jeans scale
\begin{equation}
    \lambda_{J,a} \sim 0.1 \text{ Mpc} \left(\frac{10^{-22}\text{ eV}}{m_a}\right)^{1/2} (1+z)^{1/4} \, .
\end{equation}
This stems from a quantum pressure effect, that prevents gravitational collapse on sufficiently small scales.
The practical effect is the appearance of a new term in the linear evolution equations above, playing the role of an ``effective sound speed'' for the axion~\cite{Marsh2015}:
\begin{subequations}
\label{MBPT set of equations for axion DM}
\begin{align}
    & \frac{\partial \delta_a}{\partial t} = \frac{i}{a}\Vec{v}_{\text{ba}}^{\text{(bg)}}\cdot \Vec{k}\delta_a -\theta_a \, , \\
    & \frac{\partial \theta_a}{\partial t} = \frac{i}{a}\Vec{v}_{\text{ba}}^{\text{(bg)}}\cdot \Vec{k}\theta_a -\frac{3H^2}{2}\left(\Omega_a\delta_a+\Omega_b\delta_b\right) -2H\theta_a +\frac{k^4}{4m_a^2a^4}\delta_a \, , \\
    & \frac{\partial \delta_b}{\partial t} = -\theta_b \, , \\
    & \frac{\partial \theta_b}{\partial t} = -\frac{3H^2}{2}\left(\Omega_a\delta_a+\Omega_b\delta_b\right) -2H\theta_b +\frac{c_s^2k^2}{a^2}\delta_b \, ,
\end{align}
\end{subequations}
with
\begin{equation}
    c_a^2 \approx \frac{k^2}{4m_a^2a^2} \, .
\end{equation}

\begin{figure}[h!]
    \centering
    \includegraphics[width=.6\textwidth]{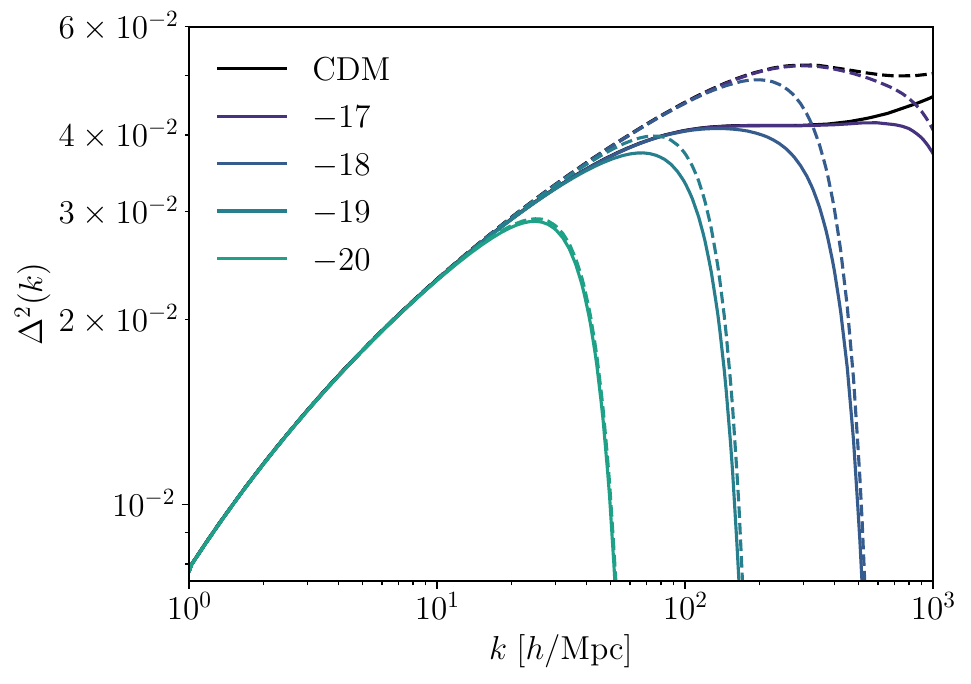}
    \caption{Effect of relative velocity and of ULA dark matter combined on the power spectrum, reproducing~\cite{Marsh2015}. Dashed lines without $\vbc$, solid lines with $\vbc$.}
    \label{fig:Pk_vbc&ULAs_z30}
\end{figure}

\begin{figure}[h!]
    \centering
    \includegraphics[width=.8\textwidth]{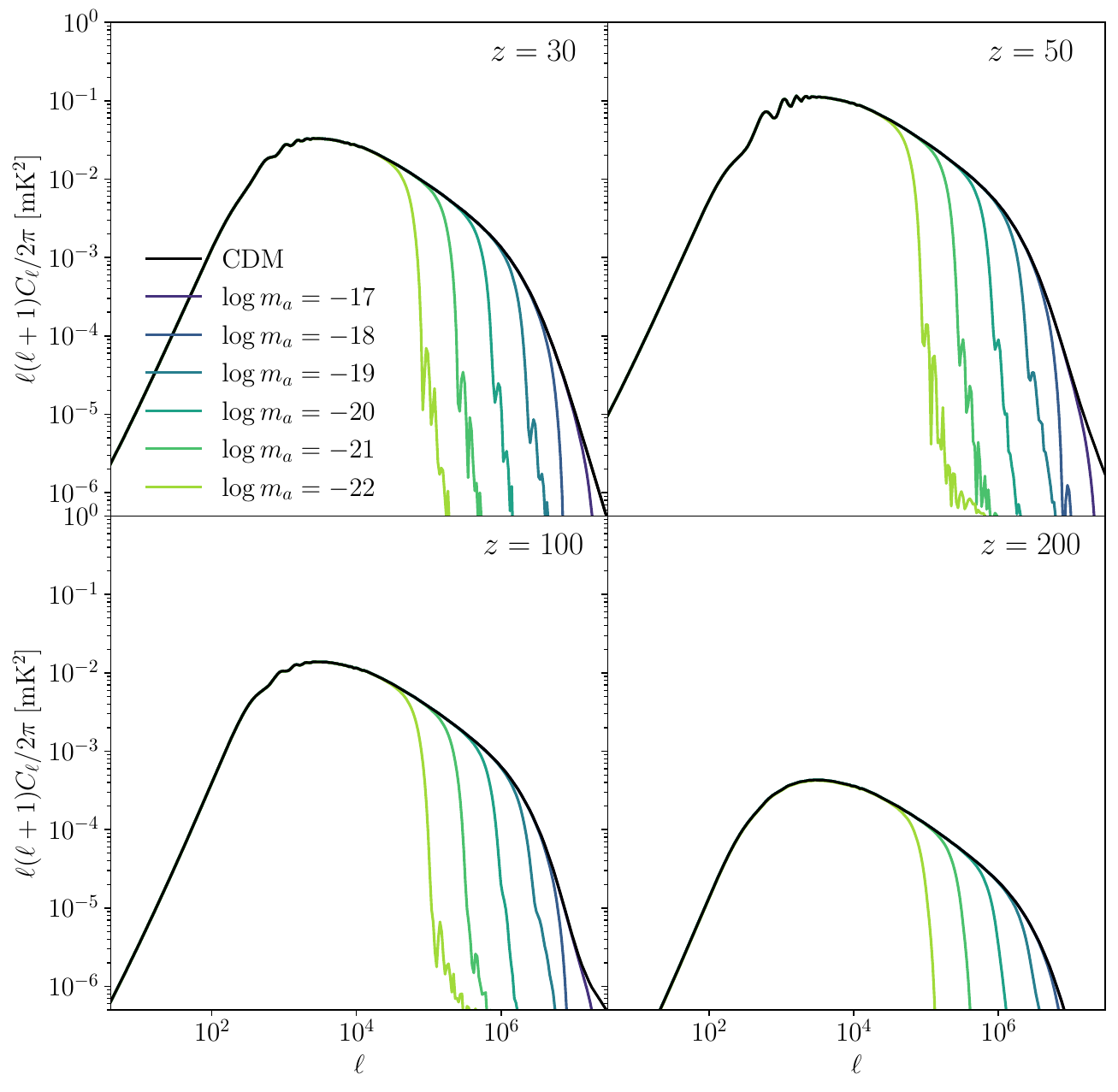}
    \caption{Effect of ULA dark matter on the 21cm angular power spectrum.}
    \label{fig:Cls_CDM&ULAs_zAll_1storder}
\end{figure}

In particular, axion dark matter with mass $m_a \lesssim 10^{-18}$ eV washes out small-scale power before the modulation effects due to relative velocity (described in the previous Section) become important. The interplay between these two effects is crucial when second order long-short mode coupling is taken into account: if the axion is light enough to wipe out power before $\vbc$ effects enter the game, then no large-scale enhancement will be present; on the contrary, for larger axion masses, the $\vbc$-induced suppression and therefore also the large-scale second order effect will be visible. Figure~\ref{fig:Pk_vbc&ULAs_z30} shows the interplay between the two effects. Figure~\ref{fig:Cls_CDM&ULAs_zAll_1storder} shows the corresponding 21cm angular power spectrum, for different values of the axion mass, compared to the CDM one.

It is also possible for axions to make up only a fraction of the total dark matter content; from now on, $\fFDM \equiv \rho_a / \rho_m$ will indicate the fraction of fuzzy dark matter, where $\rho_a$ and $\rho_m$ are the (background) densities of axions and total dark matter respectively (after the axion begins oscillating and behaves as matter, this ratio can be evaluated at whatever epoch).
For the purpose of this work, it is meaningful to focus on masses that are relevant for the interplay with the $\vbc$ effect.

In this case, the full set of equations becomes:
\begin{align}
    & \frac{\partial\delta_c}{\partial t} +\frac{1}{a}\Vec{v}_c\cdot\nabla\delta_c = -\frac{1}{a}\left(1+\delta_c\right)\nabla\cdot\Vec{v}_c \, , \\
    & \frac{\partial\Vec{v}_c}{\partial t} +\frac{1}{a}\left(\Vec{v}_c\cdot\nabla\right)\Vec{v}_c = -\frac{1}{a}\nabla\Phi -H\Vec{v}_c \, , \\
    & \frac{\partial\delta_a}{\partial t} +\frac{1}{a}\Vec{v}_a\cdot\nabla\delta_c = -\frac{1}{a}\left(1+\delta_a\right)\nabla\cdot\Vec{v}_a \, , \\
    & \frac{\partial\Vec{v}_a}{\partial t} +\frac{1}{a}\left(\Vec{v}_a\cdot\nabla\right)\Vec{v}_a = -\frac{1}{a}\nabla\left(\Phi+Q\right) -H\Vec{v}_a \, , \\
    & \frac{\partial\delta_b}{\partial t} +\frac{1}{a}\Vec{v}_b\cdot\nabla\delta_b = -\frac{1}{a}\left(1+\delta_b\right)\nabla\cdot\Vec{v}_b \, , \\
    & \frac{\partial\Vec{v}_b}{\partial t} +\frac{1}{a}\left(\Vec{v}_b\cdot\nabla\right)\Vec{v}_b = -\frac{1}{a}\nabla\Phi -H\Vec{v}_b -\frac{1}{a}c_s^2\nabla\delta_b \, , \\
    & \nabla^2\Phi = 4\pi G a^2 \Bar{\rho}_m\delta_m \, ,
\end{align}
with $Q$ the quantum potential for the axion~\cite{Marsh2015}, and:
\begin{equation}
    \delta_m = (1-\fFDM)\delta_c +\fFDM\delta_a \, .
\end{equation}

\begin{figure}[h!]
	\centering
        \includegraphics[width=.6\textwidth]{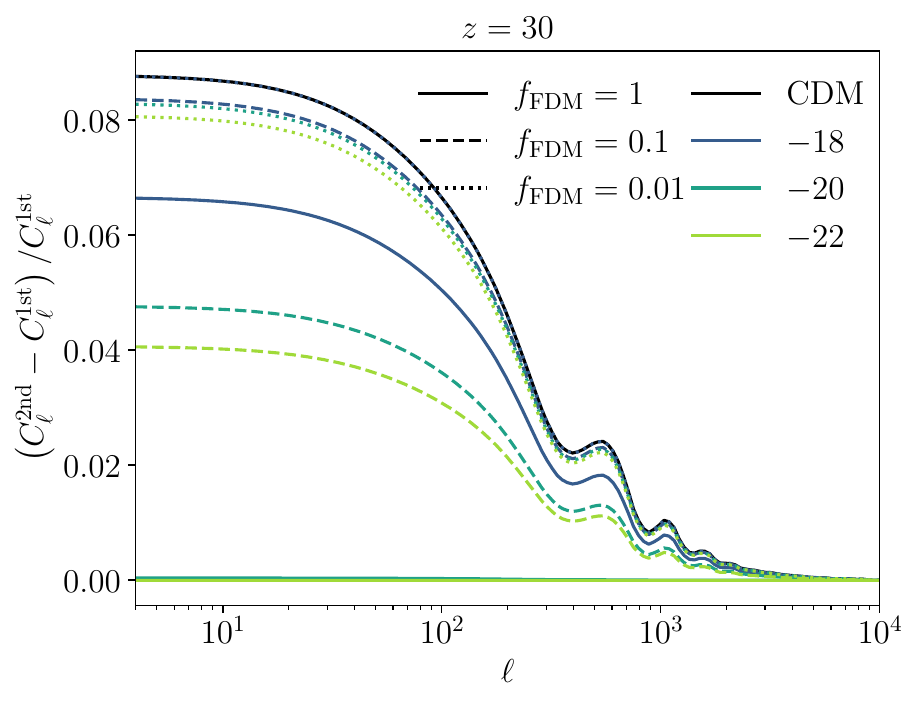}
        \caption{Second order enhancement for CDM and ULAs, for different masses and for different values of $\fFDM$.}
        \label{fig:Cls_RelCorr_comparisons1}
\end{figure}

\begin{figure}[h!]
	\centering
        \includegraphics[width=.6\textwidth]{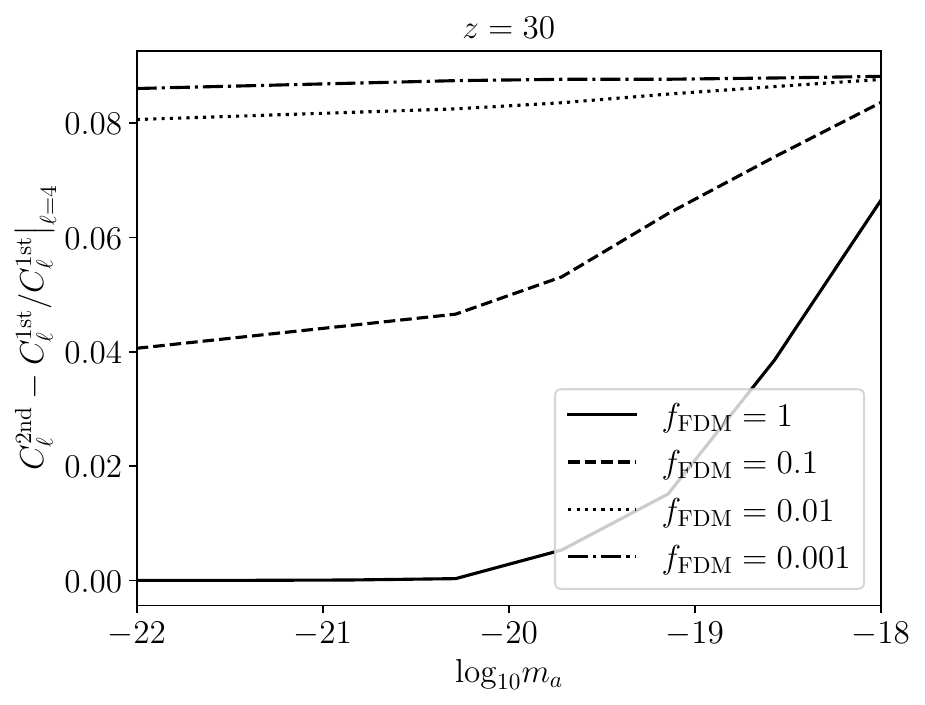} 
        \caption{Relative correction for different masses and different values of $\fFDM$, plotted at fixed $\ell=4$.}
        \label{fig:Cls_RelCorr_comparisons2}
\end{figure}

\begin{figure}[h!]
    \centering
    \includegraphics[width=.6\textwidth]{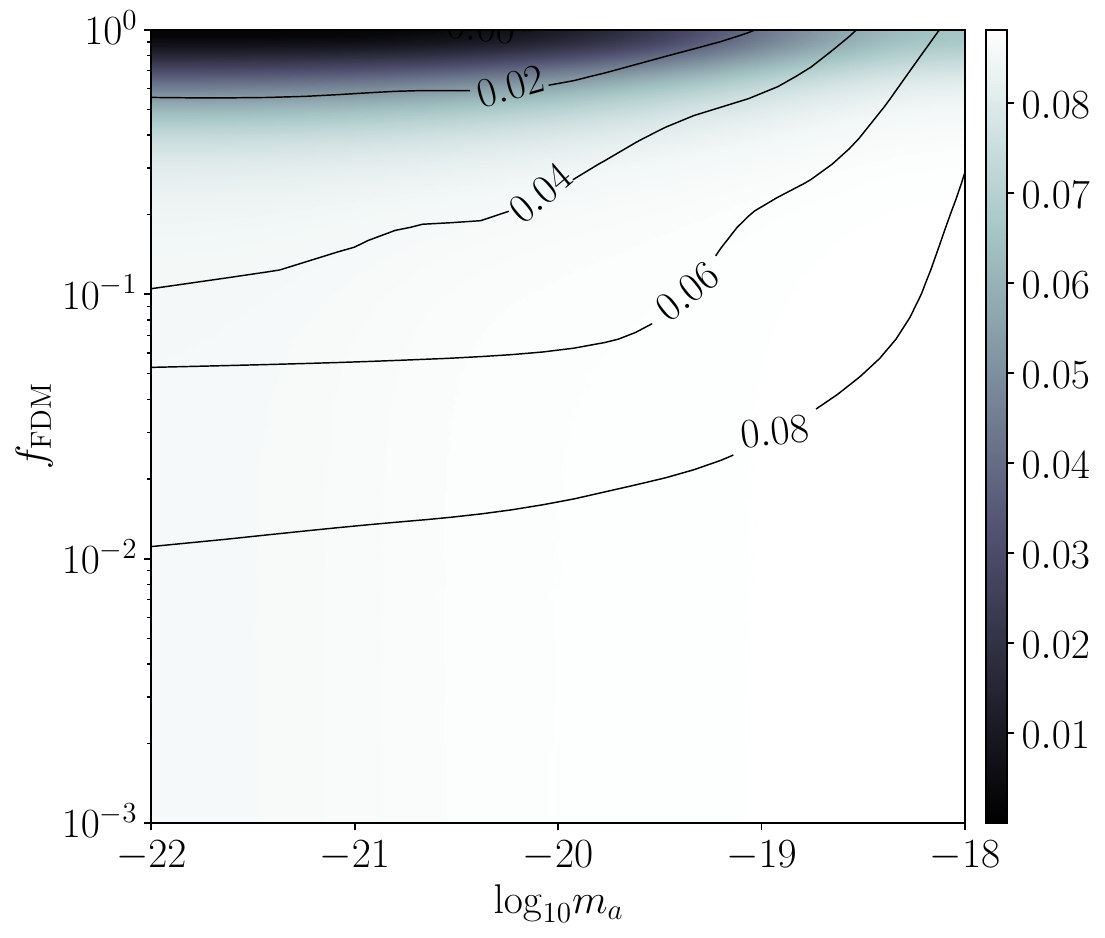}
    \caption{Ratio $\left( C_{\ell}^{\rm 2nd}-C_{\ell}^{\rm 1st} \right) / C_{\ell}^{\rm 1st}$, with varying mass and varying FDM fraction.}
    \label{fig:Cls_RelCorr_comparisons3}
\end{figure}

Figures~\ref{fig:Cls_RelCorr_comparisons1}-\ref{fig:Cls_RelCorr_comparisons2}-\ref{fig:Cls_RelCorr_comparisons3} show a comparison between second order effects for CDM and ULAs, for different $\fFDM$. The effects are plotted at $z=30$, where they are expected to be more appreciable (see~\cite{Ali-Haimoud2013} for details).

ULAs behave as CDM on large enough scales, therefore in this regime axions and cold dark matter travel with the same velocity relative to the baryons. This holds true up to the scale where quantum potential effects become important. Axions have a scale-dependent effective sound speed $c_a^2$, that could slow down axions with respect to the cold dark matter component, and bring the axion velocity field closer to that of baryons, thus suppressing the relative velocity effect and its implications from a certain $k$ onwards.
In order to apply MBPT as before, one would need to know the hierarchy of scales, which depends on the axion mass. However such a detailed analysis would be frustrated by all the simplifying assumptions that were made to write down the axion hydrodynamic equations, and this investigation is left for a future work.

\section{Analysis}
\label{sec:analysis}
Taking into account all the effects discussed in the previous sections, one can make predictions for the observational consequences of the existence of a non-negligible fraction of the dark matter in ultra-light axions.

To summarize, ULAs would cause a small-scale suppression of power. Over a specific range of scales, the baryon-dark matter relative velocity causes a dip in the matter power spectrum that in turn generates a large-scale enhancement due to long-short mode coupling.
If the suppression due to the presence of ULAs is happening on scales large enough to cut the power spectrum and therefore avoid the large-scale enhancement, low-$\ell$ measurements of the power spectrum would be able to infer the existence of the small-scale suppression.

At the linear level, in order to appreciate the small-scale suppression imprinted by the axion Jeans scale, one would need to measure the angular power spectrum up to very high $\ell \sim 10^6-10^7$, depending on the mass scale of the axion; this is futuristic even for lunar-based instruments.
However, the second order effect mentioned above, depending on the interplay between axion-induced and $\vbc$-induced suppression, could cause a detectable change in the large-scale power spectrum as measured by more realistic instruments.

Taking advantage of the separation of scales, the gravitational collapse is treated in two different regimes: on small scales ($k>1$ Mpc/$h$) through the MBPT system of coupled differential equations above, both for the CDM, ULAs, and mixed cases, and then averaging over the relative velocity; on large scales, with the Boltzmann code CLASS~\cite{Blas2011}. Initial conditions on the transfer functions are set using CLASS at decoupling. This yields all the transfer functions, for $\left\{ \delta_b, \theta_b, \delta_c, \theta_c, \delta_a, \theta_a, \delta_{T_{\rm gas}}, \delta_{x_e} \right\}$, which are needed to deal with 21cm dark ages physics. Then the 21cm angular power spectrum for temperature fluctuations is computed, switching on the Limber approximation at $\ell \sim (1\text{ $h$/Mpc}) / \chi(z)$ corresponding roughly to $\ell \sim 10^4$. The $C_{\ell}$s are computed both with and without the second order effects at low multipoles.

\subsection{Detectability}
\label{sec:fisher}
In order to establish whether a given setup will be able to detect (or rule out) certain masses and fractions of ULAs, one can compute the signal-to-noise ratio (SNR) by taking the logarithm of the likelihood to find the $\Delta\chi^{(2)}$, as~\cite{Scelfo2018}:
\begin{equation}
    \text{SNR}^2 \sim f_{\rm sky} \sum_z \sum_{\ell} \frac{2\ell+1}{2} \left( \frac{C_{\ell}^{\rm ULAs}(z) -C_{\ell}^{\rm CDM}(z)}{C^{\rm th+noise}_{\ell}(z)} \right)^2 \, .
\end{equation}
The sum runs up to an $\ell_{\rm max}$ that is taken to be the largest achievable coverage $\ell_{\rm cover}(z_{\rm min})$. The maximum achievable $\ell$ depends on the redshift, and is computed in each redshift bin as $\ell_{\rm cover}(z)$. For a given $z$, the $C_{\ell}(z)$ vector will only contain noise for $\ell > \ell_{\rm cover}(z)$

Then, at fixed axion masses, a Fisher forecast can allow to understand how well the axion mass will be constrained, marginalizing over cosmological parameters $\{ \omega_{\rm b},\omega_{\rm cdm},h,n_s,\ln 10^{10}A_s \}$. The Fisher analysis follows the formalism outlined in~\cite{Bellomo2020,Scelfo2021}.
If the maximum likelihood estimation can be well approximated by a multivariate Gaussian, then the Fisher matrix is:
\begin{equation}
    F_{\alpha\beta} = \sum_{\ell} \sigma_{\ell}^{-2} \frac{\partial C_{\ell}}{\partial\theta_{\alpha}} \frac{\partial C_{\ell}}{\partial\theta_{\beta}} \, ,
\end{equation}
where
\begin{equation}
 \sigma_{\ell}^2 = \frac{2(C^{\rm th+noise}_{\ell})^2}{f_{\rm sky}(2\ell+1)} \, ,
\end{equation}
and again the sum runs up to $\ell_{\rm max}$, considering only $z$-bins auto-correlations, as the cross-bin signal is very small for 21cm IM measurements~\cite{Hall2013}.

The noise power spectrum is diagonal if one assumes that the noises between $i$-th and $j$-th frequency channels are uncorrelated.

\subsection{Survey specifications}
The setups considered in our analysis are listed in Table~\ref{tab:survey_specs}. This paper focuses on a planned version of the SKAO, a hypothetical future extension of it (advanced SKAO, aSKAO), optimistically assumed to reach up to $z < 35$, and two possible configurations of a Lunar Radio Array (LRA) on the far side of the Moon.
LRAI and LRAII are two ideal instruments on the far side of the Moon, the former with a baseline of 1000 km, the latter covering the entire far side Moon, shown here just for the sake of indicating what could be in principle achieved.

\begin{table}[]
    \centering
    \begin{tabular}{ll|cccc}
        & & SKAO & aSKAO & LRAI & LRAII \\
        \hline
        $B$ & [MHz] & 1 & 1 & 1 & 1 \\
        $D_{\rm base}$ & [km] & 6 & 100 & 1000 & 3474 \\
        $f_{\rm cover}$ & & 0.02 & 0.2 & 0.9 & 0.9 \\
        $N_{\rm years}$ & & 5 & 10 & 10 & 10 \\
        $f_{\rm sky}$ & & 0.75 & 0.75 & 0.75 & 0.75 \\
        \hline
        $\ell_{\rm cover}$ & & 5679 & 94682 & 946832 & 3289297 \\
        noise & [mK$^2$] & $\sim 5 \cdot 10^{-3}$ & $\sim 9 \cdot 10^{-8}$ & $\sim 5 \cdot 10^{-11}$ & $\sim 4 \cdot 10^{-12}$
    \end{tabular}
    \caption{Survey specifications.}
    \label{tab:survey_specs}
\end{table}

The noise is modelled as in~\cite{Shiraishi2016}:
\begin{equation}
    C_{\ell}^{\rm noise} = (2\pi)^3 \frac{T_{\rm sys}^2(\nu)}{B \,  t_{\rm obs} f_{\rm cover}^2} \left(\frac{1}{\ell_{\rm cover}(\nu)}\right)^2 \, ,
\end{equation}
with $B$ the bandwidth of the survey, $t_{\rm obs}$ the total time of observation, $f_{\rm cover}$ the coverage fraction, $\ell_{\rm cover} = 2\pi D_{\rm base} / \lambda(z)$ the maximum observable multipole. The system temperature is taken to be the synchrotron temperature of the observed sky:
\begin{equation}
    T_{\rm sys}(\nu) = 180 \left(\frac{180\text{ MHz}}{\nu}\right)^{2.6} \, .
\end{equation}
Notice that in general one should also correct for the instrument temperature, as in~\cite{Scelfo2021}, but since here measurements are performed at high redshift, that correction is subdominant.

The binning in redshift is set following~\cite{Munoz2015}.
The correlation length in $r$ is defined as the radial separation beyond which the cross-correlation between two redshift slices is less than 1/2 the power spectrum. Then it is converted into a correlation length in frequency:
\begin{equation}
    \xi_{\nu} \approx 1 \text{ MHz} \left(\frac{51}{1+z}\right)^{1/2} \left(\frac{\xi_r}{60 \text{ Mpc}}\right) \, ,
\end{equation}
where $\xi_r$ is the correlation length in radial comoving separation, and it can be taken to be $\xi_r \sim 60$ Mpc as a reference value. Here $\xi_{\nu}$ needs to be smaller than the bandwith $B$ of the instrument, so one can take $(\nu_{\rm max}-\nu_{\rm min})/B$ linearly spaced bins in frequency, and then convert them into redshift bins: the resulting bins will be more finely spaced at low redshifts, and broader at high redshifts.

\section{Results and discussion}
\label{sec:results}
This Section presents results for the predicted limits on axion mass from 21cm IM from different future interferometers, obtained using the effects described in the previous Section.

One can already draw some conclusions, based purely on physical and instrumental considerations, which will then be confirmed in a quantitative way with numerical results.
When looking at large scale effects, axions with $m_a \sim 10^{-18}$ eV and heavier behave as CDM, since they suppress the matter power spectrum at smaller scales than the ones over which the $\vbc$ induced suppression is visible. Lighter masses, instead, wash out the $\vbc$ features and consequently also the low-$\ell$ enhancement, providing a different scenario from standard CDM. Indeed, the SNR is higher for lighter masses. For such light values, where the difference between the two models is the most appreciable, even the ground-based aSKAO reaches large SNR values. Proposed Moon-based instruments benefit from taking into account second order effects in the analysis, but the improvement is mass-dependent and less drastic, because such surveys can take advantage of the high multipoles range they can access, in order to probe quantum pressure effects.

\begin{figure}[h!]
    \centering
    \includegraphics[width=.7\textwidth]{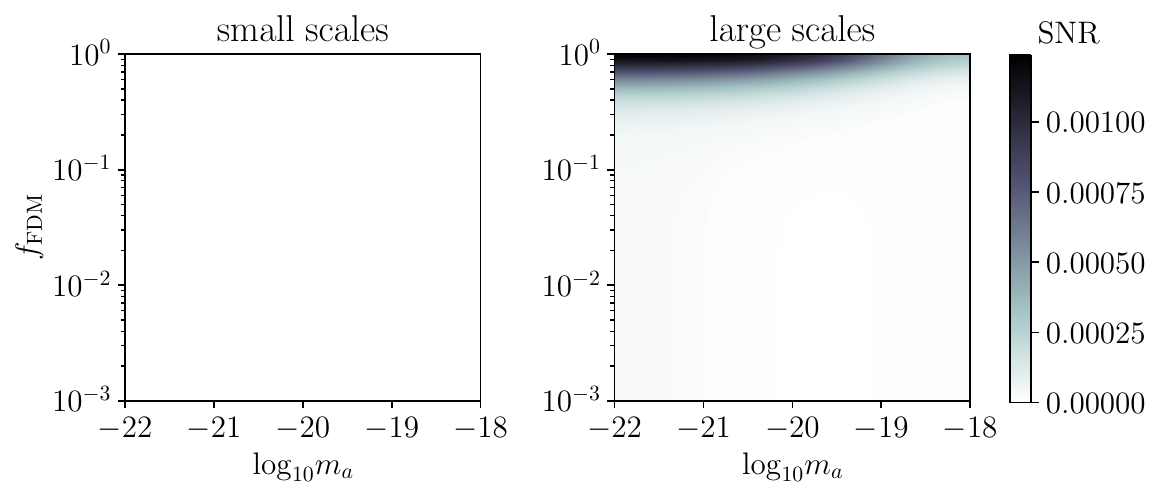}
    \caption{SNR for the small and large scale effects, for the range of masses considered in this work as a function of different fractions. Results are for the SKAO.}
    \label{fig:SNRSKAO}
\end{figure}

First of all, considering the SKAO configuration as described in Table~\ref{tab:survey_specs}, one can work with a single bin at $z \sim 30$.
The analysis (with results shown in Figure~\ref{fig:SNRSKAO}) confirms that the noise will dominate over the signal for the mass range considered here, as expected.
This is valid for both the small and large scale effects, for which the SNR reaches at best (for the smallest masses) $\sim 10^{-3}$.

If one considers the more advanced configuration aSKAO, in the optimistic redshift range $z \sim 30-35$, detection (or ruling out) of most masses considered would become possible.

\begin{figure}[h!]
    \centering
    \includegraphics[width=.7\textwidth]{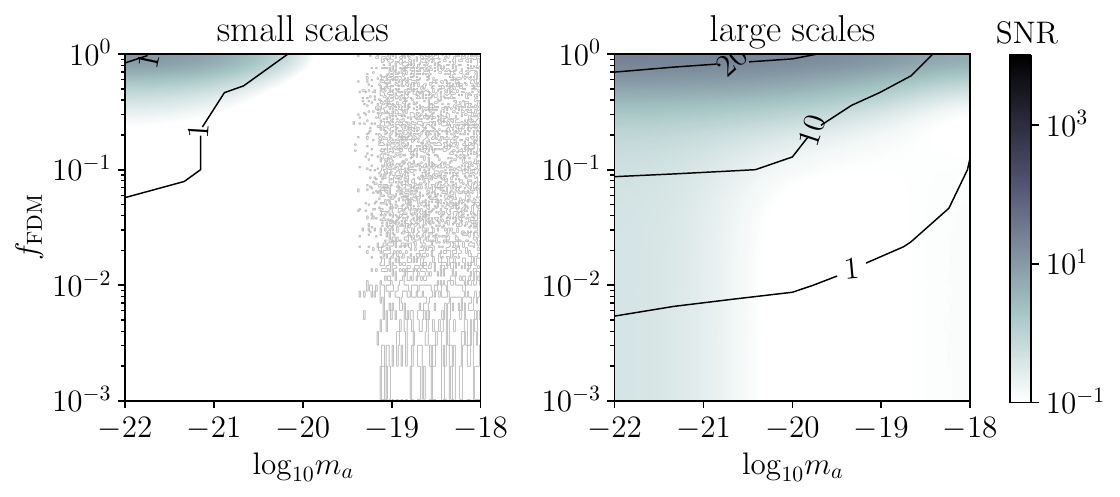}
    \caption{Same as Figure~\ref{fig:SNRSKAO} but for aSKAO.}
    \label{fig:SNRaSKAO}
\end{figure}

Looking more in detail (see Figure~\ref{fig:SNRaSKAO}), when considering only small-scale effects, even in this configuration one could have a ${\rm SNR}>1$ only for the smallest masses considered. This is because ground based realistic multipole ranges (which are proportional to the baseline) allow to only barely touch the beginning of the power spectrum damping from axions of masses of $\sim 10^{-21}$.
However, to confirm the power of analyses taking advantage of the long-short mode coupling, the right panel of Figure~\ref{fig:SNRaSKAO} illustrates how an upgraded version of the SKAO could reach SNR values $>10$ for even dark matter fractions of $10\%$ for $m_a \le 10^{-20}$, being able to probe larger fractions when considering heavier masses.

\begin{figure}[h!]
    \centering
    \includegraphics[width=.7\textwidth]{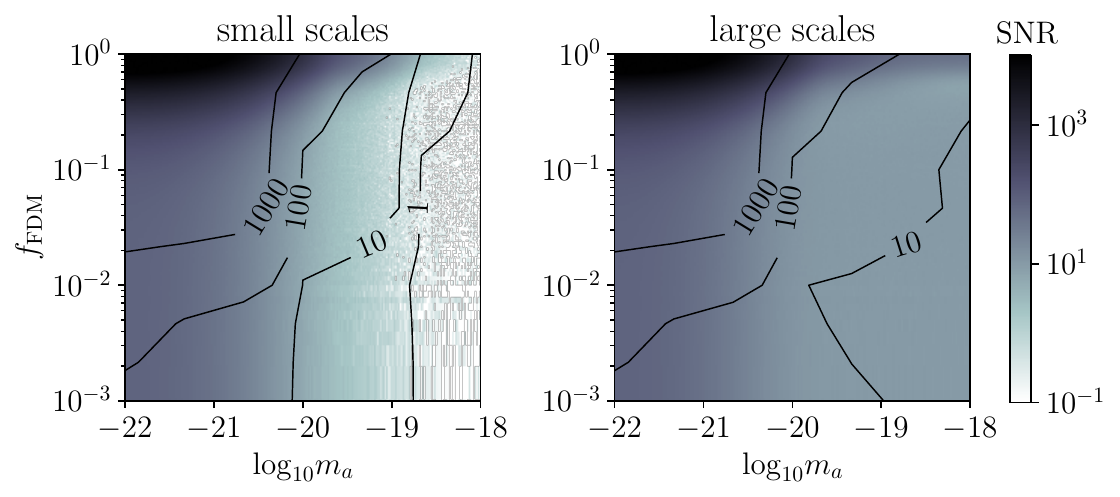}
    \caption{Same as Figure~\ref{fig:SNRSKAO} but for LRAI.}
    \label{fig:SNRLRAI}
\end{figure}

Future interferometers on the far side of the Moon will reach extremely high values of $\ell_{\rm max}$; this will allow to tap into very small scales information, unaccessible to any other observable. Therefore (see Figure~\ref{fig:SNRLRAI}), small scales will already provide very high SNR for masses $\lesssim 10^{-19}$; for this mass range, one could reach a ${\rm SNR}>10$ even for very small $\fFDM$, making those instruments able to detect virtually any (non negligible) fraction of fuzzy dark matter for $m_a \le 10^{-19}$. The sharp reduction of SNR when going to larger masses is easily understood by looking at the multipole coverage of the instrument.

In Figure~\ref{fig:SNRLRAI} one can see that for lighter masses, large scale effects only marginally improve constraints; this is again understandable, as the large $\ell$ reach of the LRAI is already powerful enough to allow very high significance detections. Going toward heavier masses, on the other hand, large scale effects open up the possibility to detect (or rule out) very small fractions of fuzzy dark matter for all considered masses. In particular, the LRAI would enable robust detections of even $0.1\%$ of dark matter in axions, for all the masses considered in this work.

\begin{figure}[h!]
    \centering
    \includegraphics[width=0.6\textwidth]{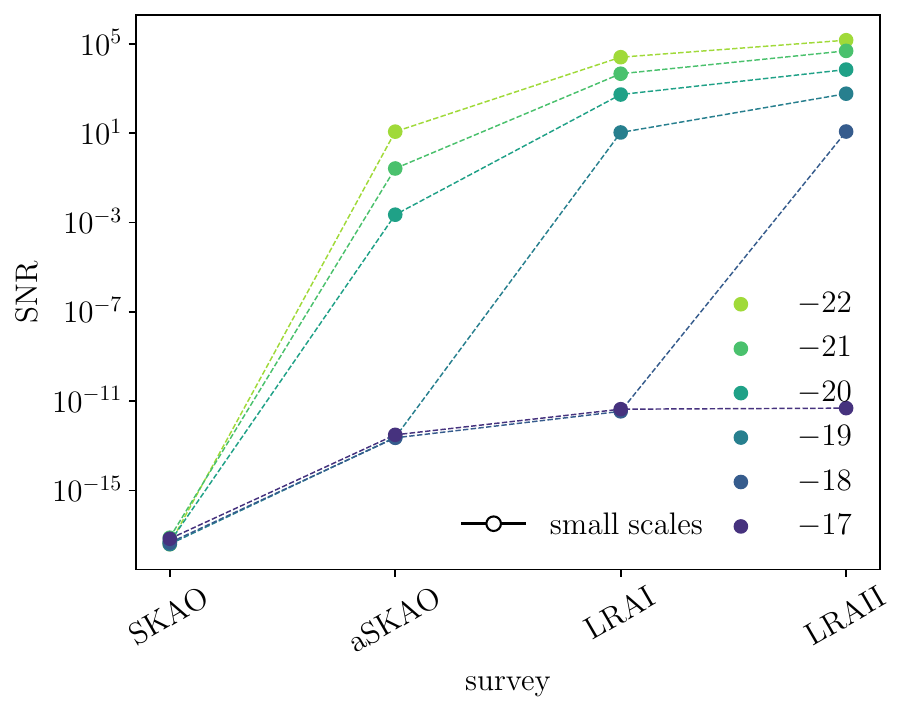}
    \includegraphics[width=0.6\textwidth]{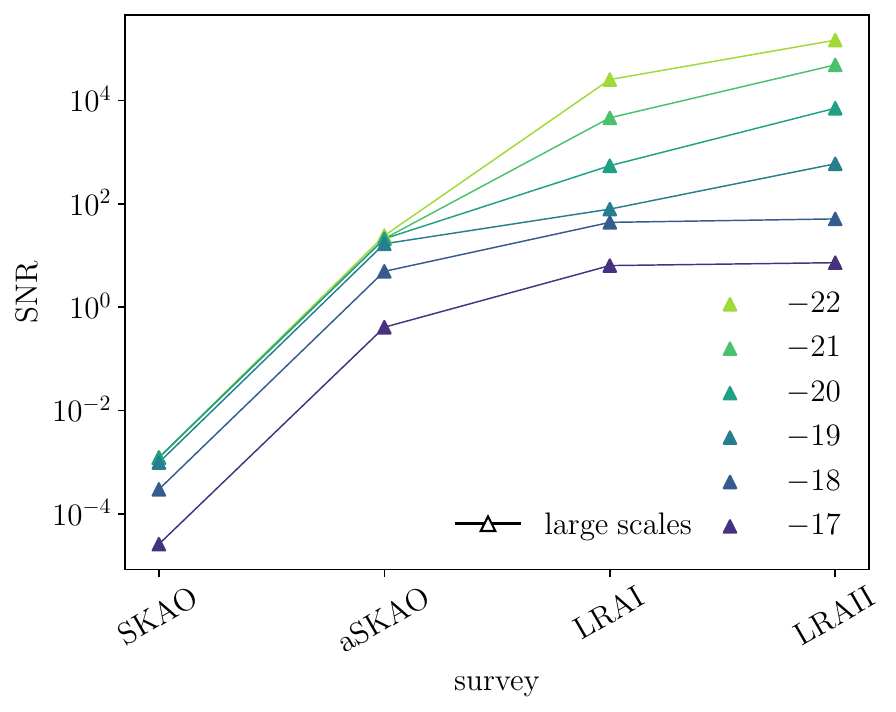}
    \caption{SNR for small  ({\it Top panel}) and large scales ({\it Bottom panel}) power spectrum modifications, for different masses and different surveys.}
\label{fig:surveys_comparison}
\end{figure}

In Figure~\ref{fig:surveys_comparison} one can find a summary of the SNR reachable, for small scale (top panel) and large scale (bottom panel) effects, for all the instruments and masses considered in this work.
This Figure also includes the hypothetical LRAII instrument, which is given as a comparison. This shows that going from a futuristic, but realistic LRAI configuration to a proof-of-principle full Moon coverage would considerably help with small-scale effects, but that the small-large scales interplay makes it that the LRAI can already reach the necessary constraining power.

In order to assess possible degeneracies with cosmological parameters, a Fisher matrix analysis is performed, over the parameters $\left\{ \omega_{\rm b}, \omega_{\rm cdm}, h, n_s, \ln10^{10}A_s\right\}$.
Our results indicate that, for experiments that are competitive for the masses considered, including variations of other cosmological parameters will not considerably change our results.

Using Fisher matrix analyses, the precision in the determination of the axion mass can be studied, assuming this will be detected; in practice, this means taking different $m_a$ as fiducial.

\begin{figure}[h!]
    \centering
    \includegraphics[width=.6\textwidth]{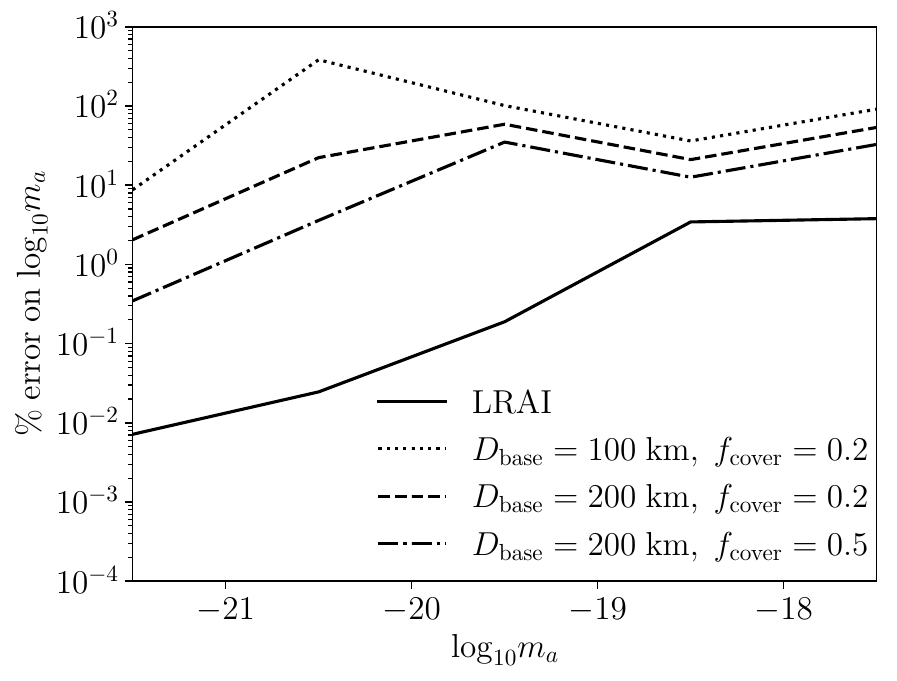}
    \caption{$1\sigma$ percentage error on $\log_{10}m_a$, including all effects (large and small scales), for three ground-based mock surveys and for the lunar-based LRAI, marginalizing over cosmological parameters, at fixed $\fFDM=1$. Constraints improve going to lighter masses, for which the improvement for a lunar array becomes larger.}
    \label{fig:1sigma_fFDM1_SKA&haSKA&LRAI}
\end{figure}

Figure~\ref{fig:1sigma_fFDM1_SKA&haSKA&LRAI} shows percentage $1\sigma$ errors for a few ground-based hypothetical surveys and for the lunar-based LRAI.
A ground-based advanced SKAO-like interferometer would be able to constraint an ULA in the mass range $10^{-21}-10^{-17}$ eV at $1\sigma \sim 30\%$, taking advantage of the large-scale effects. Building an instrument on the Moon with an LRAI-like survey would achieve $1\sigma < 10\%$ for all masses considered, and up to $\lesssim 0.01\%$ for the lightest axions.
All these constraints are for the totality of the dark matter in axions ($\fFDM=1$).

As expected, ground instrument constraints worsen with lighter masses, where no large-scale effects are expected (cf. Figure~\ref{fig:Cls_RelCorr_comparisons3}). 
While being obviously less powerful than large lunar arrays, they are most competitive around $10^{-19}-10^{-18}$ eV, but then lose constraining power as the power spectrum converges to the CDM one. A lunar-based interferometer, instead, would be able to measure the lighter axions up to $\lesssim 0.01\%$ level.

Results presented in this Section are all calculated as direct modifications of angular power spectra. A real data analysis would need to account for observational considerations including foregrounds and systematic effects. The Galactic synchrotron foreground is orders of magnitude brighter than the 21cm signal, and foreground characterization and subtraction is an open issue, see e.g.,~\cite{Zaldarriaga2003,Furlanetto2006,Furlanetto2019a}. Therefore, here only cosmic variance and instrumental noise are included, while a more detailed observational investigation is left to a future work.

Additional complications may enter the game at small scales, such as model-dependent self-interactions, and the formation of soliton cores at the center of dark matter halos~\cite{Schive2014,MarshPop2015}. The soliton profile can be modelled as~\cite{Schive2014} $\rho(r) = \rho_0 \left[ 1 +0.091 \left( r/r_c \right)^2 \right]^{-8}$, where $\rho_0$ is the central density and $r_c$ the radius at which the density reaches half of its central value,
\begin{equation}
	r_c = 1.0 \left( \frac{\rho_0}{3.1 \cdot 10^{15} \ M_{\odot}/{\rm Mpc}^3 }\right)^{1/4} \left( \frac{m_a}{2.5 \cdot 10^{-22} \ {\rm eV}} \right)^{1/2} \ \text{kpc} \, .
\end{equation}
If one takes $r_c$ as an estimate of the scale associated with the soliton core, the corresponding angular scales $\ell_c(z) \sim k_c r(z)$ are of order
\begin{equation}
	\ell_c(z) \simeq 1.1 \cdot 10^7 \left( \frac{\rho_0}{ 3.1 \cdot 10^{15} \ M_{\odot}/{\rm Mpc}^3 } \right)^{-1/4} \left( \frac{m_a}{10^{-20} \ {\rm eV}} \right)^{-1/2} \left( \frac{ r(z) }{ r(z=30) } \right) \, .
\end{equation}
For the heavier masses considered, $m_a \simeq 10^{-18} \text{ eV}$, these scales may be accessible by the most futuristic configuration LRAII, for which $\ell_{\rm cover}$ is of order $10^6$. A proper treatment would require numerical simulations, similar to what has been implemented recently in~\cite{Mocz2023,Lague2023}.

\subsection{Ground-based survey design for $\fFDM<10\%$}
\label{sec:surveyfdm}
In this Section is presented an investigation on what are the instrument requirements (for ground-based interferometers) that would allow to detect a fuzzy dark matter presence of order $\fFDM \sim \mathcal{O}(10\%)$.

\begin{figure}[h!]
    \centering
    \includegraphics[width=.7\textwidth]{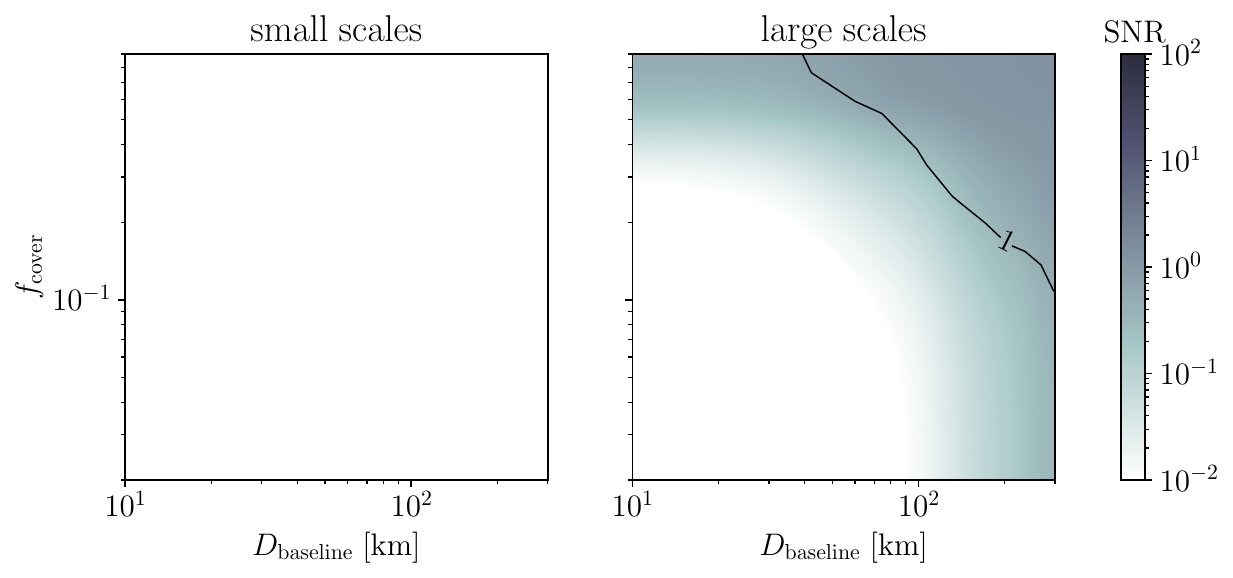}
    \includegraphics[width=.7\textwidth]{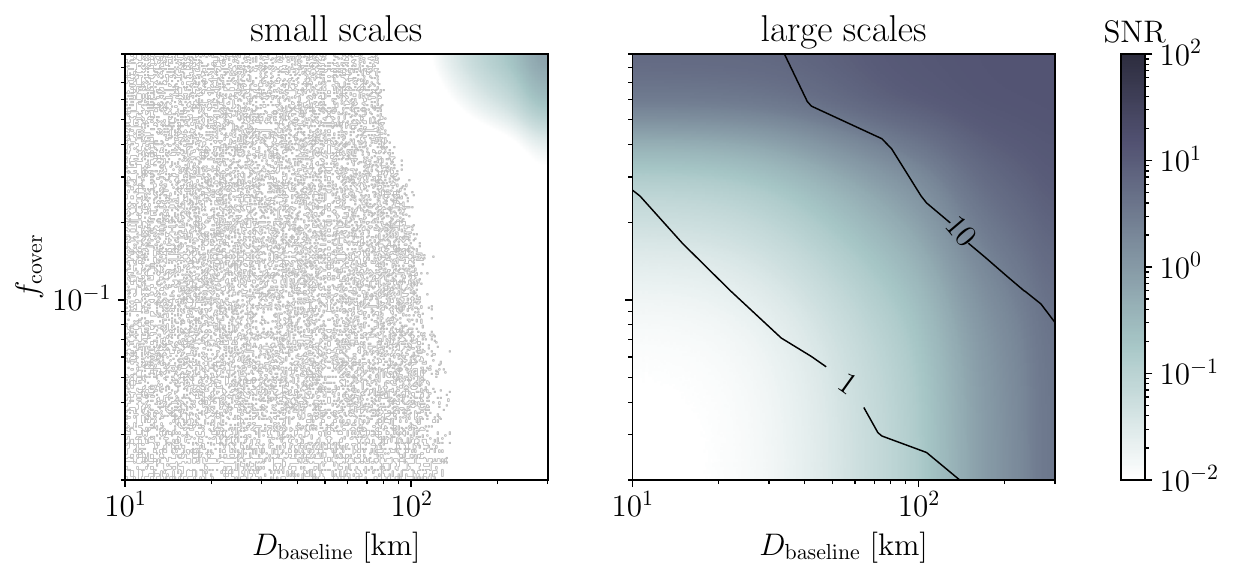}
    \includegraphics[width=.7\textwidth]{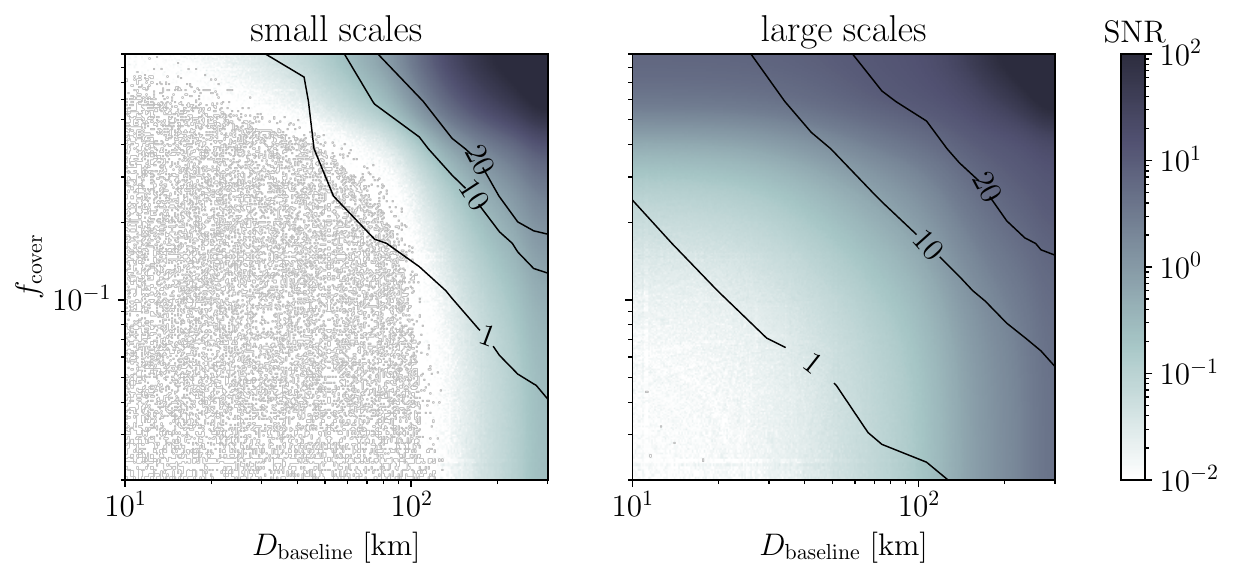}
    \caption{SNR without and with second order effects, for $\log_{10}m_a=-18,-20,-22$ (top to bottom), varying the design of the survey. Here the observation time is fixed to 10 years, $B=1$ MHz, $f_{\rm sky}=0.75$.}
    \label{fig:ContourPlotSNR_haSKA}
\end{figure}

Figure~\ref{fig:ContourPlotSNR_haSKA} shows the expected SNR when varying the baseline $D_{\rm base}$ and the area coverage $f_{\rm cover}$ of the instrument, for different axion masses.
If the axion is heavier than $\sim 10^{-20}$ eV, small-scale suppression happens at wavenumbers larger than where the $\vbc$ has relevant effects: this implies that there is no difference at low multipoles between CDM and ULA spectra. In this case, a detection must rely on the observation of higher multipoles, which are practically unaccessible for a ground based detector. On the contrary, if the axion is lighter, then increasing the baseline of the survey and/or the coverage fraction could lead to a detection.

\section{Conclusions}
\label{sec:conclusions}
Measurements of 21cm line intensity mapping power spectrum during the dark ages have the potential to unveil a wealth of cosmological information; this paper focuses on shedding light on the nature of dark matter. This work forecasts the ability to distinguish between the standard CDM scenario and a compelling alternative, namely ultra-light axion-like scalar fields; these have the peculiarity of suppressing the matter power spectrum on small scales, due to quantum pressure effects.
Ultra-light axion dark matter has become one of the most promising candidates in the recent years, and many works set (sometimes tentative) constraints on its abundance and mass. Among the models considered in literature, masses of around $10^{-22} - 10^{-18}$ eV are currently the least robustly constrained.

This study considers future planned surveys, like an advanced version of the Square Kilometer Array Observatory, and a proposed lunar-based radio array, which would gain access to the full dark ages. 

The relative velocity $\vbc$ between baryons and dark matter at recombination introduces a suppression of the power spectrum over a range of specific scales ($\approx 50 - 1000 \ h{\rm /Mpc}$).
This relative velocity induces a long-short mode coupling which leads to a large-scale enhancement in the power spectrum, due to the fact that the $\vbc$ has a coherence length that is much smaller than the cosmological horizon.

Depending on the axion mass, this feature may or may not be washed out by the axion-induced suppression, which happens at scales that are directly related to the axion mass.

By forecasting the multipoles reach and the precision in measuring the 21cm IM spectrum for different proposed radio interferometers, this work studies their capabilities to constrain the presence of ultra-light axions of different masses and abundances.

The results obtained by using both the small-scale suppression and the large-scale enhancement show how currently proposed ground-based instruments will not be able to set meaningful constraints for the axion masses considered here.
On the other hand, possible future extensions of ground-based interferometers could be able to help setting strong constraints on the masses mentioned above.

Building a radio array on the far, radio-silent side of the Moon would allow to properly tap into the dark ages and both measure very precisely the large-scale 21cm power spectrum and reach extremely high multipoles.
This would allow to distinguish ULAs from CDM with high significance, even if axions were to make up as low as $1\%$ of the dark matter, and essentially close the axion dark matter window on currently allowed masses.

\begin{acknowledgments}
We thank Jose Luis Bernal, Kimberly Boddy and Ely Kovetz for insightful discussions, and Sarah Libanore, Alice Cipolla and Michele Sanna for helpful comments.
AR acknowledges funding from the Italian Ministry of University and Research (MIUR) through the ``Dipartimenti di eccellenza'' project ``Science of the Universe''.
This work is supported in part by the MUR Departments of Excellence grant ``Quantum Frontiers''.
\end{acknowledgments}

\appendix



\bibliography{bibliography.bib}

\providecommand{\href}[2]{#2}\begingroup\raggedright\begin{thebibliography}{10}

\bibitem{Bertone2004}
G.~Bertone, D.~Hooper, and J.~Silk, ``{Particle dark matter: Evidence,
  candidates and constraints}'',
  \href{http://dx.doi.org/10.1016/j.physrep.2004.08.031}{{\em Phys. Rept.}
  {\bfseries 405} (2005) 279--390},
  \href{http://arxiv.org/abs/hep-ph/0404175}{{\ttfamily arXiv:hep-ph/0404175}}.

\bibitem{Bullock2017}
J.~S. Bullock and M.~Boylan-Kolchin, ``{Small-Scale Challenges to the
  $\Lambda$CDM Paradigm}'',
  \href{http://dx.doi.org/10.1146/annurev-astro-091916-055313}{{\em Ann. Rev.
  Astron. Astrophys.} {\bfseries 55} (2017) 343--387},
  \href{http://arxiv.org/abs/1707.04256}{{\ttfamily arXiv:1707.04256
  [astro-ph.CO]}}.

\bibitem{Hu2000}
W.~Hu, R.~Barkana, and A.~Gruzinov, ``{Cold and fuzzy dark matter}'',
  \href{http://dx.doi.org/10.1103/PhysRevLett.85.1158}{{\em Phys. Rev. Lett.}
  {\bfseries 85} (2000) 1158--1161},
  \href{http://arxiv.org/abs/astro-ph/0003365}{{\ttfamily
  arXiv:astro-ph/0003365}}.

\bibitem{Marsh2015review}
D.~J.~E. Marsh, ``{Axion Cosmology}'',
  \href{http://dx.doi.org/10.1016/j.physrep.2016.06.005}{{\em Phys. Rept.}
  {\bfseries 643} (2016) 1--79},
  \href{http://arxiv.org/abs/1510.07633}{{\ttfamily arXiv:1510.07633
  [astro-ph.CO]}}.

\bibitem{Hui2016}
L.~Hui, J.~P. Ostriker, S.~Tremaine, and E.~Witten, ``{Ultralight scalars as
  cosmological dark matter}'',
  \href{http://dx.doi.org/10.1103/PhysRevD.95.043541}{{\em Phys. Rev. D}
  {\bfseries 95} no.~4, (2017) 043541},
  \href{http://arxiv.org/abs/1610.08297}{{\ttfamily arXiv:1610.08297
  [astro-ph.CO]}}.

\bibitem{Arvanitaki2009}
A.~Arvanitaki, S.~Dimopoulos, S.~Dubovsky, N.~Kaloper, and J.~March-Russell,
  ``{String Axiverse}'',
  \href{http://dx.doi.org/10.1103/PhysRevD.81.123530}{{\em Phys. Rev. D}
  {\bfseries 81} (2010) 123530},
  \href{http://arxiv.org/abs/0905.4720}{{\ttfamily arXiv:0905.4720 [hep-th]}}.

\bibitem{Kamionkowski2014}
M.~Kamionkowski, J.~Pradler, and D.~G.~E. Walker, ``{Dark energy from the
  string axiverse}'',
  \href{http://dx.doi.org/10.1103/PhysRevLett.113.251302}{{\em Phys. Rev.
  Lett.} {\bfseries 113} no.~25, (2014) 251302},
  \href{http://arxiv.org/abs/1409.0549}{{\ttfamily arXiv:1409.0549 [hep-ph]}}.

\bibitem{Visinelli2018}
L.~Visinelli and S.~Vagnozzi, ``{Cosmological window onto the string axiverse
  and the supersymmetry breaking scale}'',
  \href{http://dx.doi.org/10.1103/PhysRevD.99.063517}{{\em Phys. Rev. D}
  {\bfseries 99} no.~6, (2019) 063517},
  \href{http://arxiv.org/abs/1809.06382}{{\ttfamily arXiv:1809.06382
  [hep-ph]}}.

\bibitem{Sarkar2022}
D.~Sarkar, J.~Flitter, and E.~D. Kovetz, ``{Exploring delaying and heating
  effects on the 21-cm signature of fuzzy dark matter}'',
  \href{http://dx.doi.org/10.1103/PhysRevD.105.103529}{{\em Phys. Rev. D}
  {\bfseries 105} no.~10, (2022) 103529},
  \href{http://arxiv.org/abs/2201.03355}{{\ttfamily arXiv:2201.03355
  [astro-ph.CO]}}.

\bibitem{Hotinli2022}
S.~C. Hotinli, D.~J.~E. Marsh, and M.~Kamionkowski, ``{Probing ultralight
  axions with the 21-cm signal during cosmic dawn}'',
  \href{http://dx.doi.org/10.1103/PhysRevD.106.043529}{{\em Phys. Rev. D}
  {\bfseries 106} no.~4, (2022) 043529},
  \href{http://arxiv.org/abs/2112.06943}{{\ttfamily arXiv:2112.06943
  [astro-ph.CO]}}.

\bibitem{Brito2015}
R.~Brito, V.~Cardoso, and P.~Pani, ``{Superradiance}: {New Frontiers in Black
  Hole Physics}'', \href{http://dx.doi.org/10.1007/978-3-319-19000-6}{{\em
  Lect. Notes Phys.} {\bfseries 906} (2015) pp.1--237},
  \href{http://arxiv.org/abs/1501.06570}{{\ttfamily arXiv:1501.06570 [gr-qc]}}.

\bibitem{Viel2013}
M.~Viel, G.~D. Becker, J.~S. Bolton, and M.~G. Haehnelt, ``{Warm dark matter as
  a solution to the small scale crisis: New constraints from high redshift
  Lyman-\ensuremath{\alpha} forest data}'',
  \href{http://dx.doi.org/10.1103/PhysRevD.88.043502}{{\em Phys. Rev. D}
  {\bfseries 88} (2013) 043502},
  \href{http://arxiv.org/abs/1306.2314}{{\ttfamily arXiv:1306.2314
  [astro-ph.CO]}}.

\bibitem{Kobayashi2017}
T.~Kobayashi, R.~Murgia, A.~De~Simone, V.~Ir\v{s}i\v{c}, and M.~Viel,
  ``{Lyman-$\alpha$ constraints on ultralight scalar dark matter: Implications
  for the early and late universe}'',
  \href{http://dx.doi.org/10.1103/PhysRevD.96.123514}{{\em Phys. Rev. D}
  {\bfseries 96} no.~12, (2017) 123514},
  \href{http://arxiv.org/abs/1708.00015}{{\ttfamily arXiv:1708.00015
  [astro-ph.CO]}}.

\bibitem{Rogers2020}
K.~K. Rogers and H.~V. Peiris, ``{Strong Bound on Canonical Ultralight Axion
  Dark Matter from the Lyman-Alpha Forest}'',
  \href{http://dx.doi.org/10.1103/PhysRevLett.126.071302}{{\em Phys. Rev.
  Lett.} {\bfseries 126} no.~7, (2021) 071302},
  \href{http://arxiv.org/abs/2007.12705}{{\ttfamily arXiv:2007.12705
  [astro-ph.CO]}}.

\bibitem{Flitter2022}
J.~Flitter and E.~D. Kovetz, ``{Closing the window on fuzzy dark matter with
  the 21-cm signal}'',
  \href{http://dx.doi.org/10.1103/PhysRevD.106.063504}{{\em Phys. Rev. D}
  {\bfseries 106} no.~6, (2022) 063504},
  \href{http://arxiv.org/abs/2207.05083}{{\ttfamily arXiv:2207.05083
  [astro-ph.CO]}}.

\bibitem{Hlozek2014}
R.~Hlozek, D.~Grin, D.~J.~E. Marsh, and P.~G. Ferreira, ``{A search for
  ultralight axions using precision cosmological data}'',
  \href{http://dx.doi.org/10.1103/PhysRevD.91.103512}{{\em Phys. Rev. D}
  {\bfseries 91} no.~10, (2015) 103512},
  \href{http://arxiv.org/abs/1410.2896}{{\ttfamily arXiv:1410.2896
  [astro-ph.CO]}}.

\bibitem{Hlozek2017}
R.~Hlozek, D.~J.~E. Marsh, and D.~Grin, ``{Using the Full Power of the Cosmic
  Microwave Background to Probe Axion Dark Matter}'',
  \href{http://dx.doi.org/10.1093/mnras/sty271}{{\em Mon. Not. Roy. Astron.
  Soc.} {\bfseries 476} no.~3, (2018) 3063--3085},
  \href{http://arxiv.org/abs/1708.05681}{{\ttfamily arXiv:1708.05681
  [astro-ph.CO]}}.

\bibitem{Acharya2010}
B.~S. Acharya, K.~Bobkov, and P.~Kumar, ``{An M Theory Solution to the Strong
  CP Problem and Constraints on the Axiverse}'',
  \href{http://dx.doi.org/10.1007/JHEP11(2010)105}{{\em JHEP} {\bfseries 11}
  (2010) 105}, \href{http://arxiv.org/abs/1004.5138}{{\ttfamily arXiv:1004.5138
  [hep-th]}}.

\bibitem{Acharya2010a}
B.~S. Acharya, G.~Kane, and E.~Kuflik, ``{Bounds on scalar masses in theories
  of moduli stabilization}'',
  \href{http://dx.doi.org/10.1142/S0217751X14500730}{{\em Int. J. Mod. Phys. A}
  {\bfseries 29} (2014) 1450073},
  \href{http://arxiv.org/abs/1006.3272}{{\ttfamily arXiv:1006.3272 [hep-ph]}}.

\bibitem{Mehta2021}
V.~M. Mehta, M.~Demirtas, C.~Long, D.~J.~E. Marsh, L.~McAllister, and M.~J.
  Stott, ``{Superradiance in string theory}'',
  \href{http://dx.doi.org/10.1088/1475-7516/2021/07/033}{{\em JCAP} {\bfseries
  07} (2021) 033}, \href{http://arxiv.org/abs/2103.06812}{{\ttfamily
  arXiv:2103.06812 [hep-th]}}.

\bibitem{Kovetz2017}
E.~D. Kovetz {\em et~al.}, ``{Line-Intensity Mapping: 2017 Status Report}'',
  \href{http://arxiv.org/abs/1709.09066}{{\ttfamily arXiv:1709.09066
  [astro-ph.CO]}}.

\bibitem{Tseliakhovich2010}
D.~Tseliakhovich and C.~Hirata, ``{Relative velocity of dark matter and
  baryonic fluids and the formation of the first structures}'',
  \href{http://dx.doi.org/10.1103/PhysRevD.82.083520}{{\em Phys. Rev. D}
  {\bfseries 82} (2010) 083520},
  \href{http://arxiv.org/abs/1005.2416}{{\ttfamily arXiv:1005.2416
  [astro-ph.CO]}}.

\bibitem{Ali-Haimoud2013}
Y.~Ali-Ha\"\i{}moud, P.~D. Meerburg, and S.~Yuan, ``{New light on 21 cm
  intensity fluctuations from the dark ages}'',
  \href{http://dx.doi.org/10.1103/PhysRevD.89.083506}{{\em Phys. Rev. D}
  {\bfseries 89} no.~8, (2014) 083506},
  \href{http://arxiv.org/abs/1312.4948}{{\ttfamily arXiv:1312.4948
  [astro-ph.CO]}}.

\bibitem{Marsh2015}
D.~J.~E. Marsh, ``{Nonlinear hydrodynamics of axion dark matter: Relative
  velocity effects and quantum forces}'',
  \href{http://dx.doi.org/10.1103/PhysRevD.91.123520}{{\em Phys. Rev. D}
  {\bfseries 91} no.~12, (2015) 123520},
  \href{http://arxiv.org/abs/1504.00308}{{\ttfamily arXiv:1504.00308
  [astro-ph.CO]}}.

\bibitem{Planck2018}
The {\bfseries Planck}, N.~Aghanim {\em et~al.}, ``{Planck 2018 results. VI.
  Cosmological parameters}'',
  \href{http://dx.doi.org/10.1051/0004-6361/201833910}{{\em Astron. Astrophys.}
  {\bfseries 641} (2020) A6}, \href{http://arxiv.org/abs/1807.06209}{{\ttfamily
  arXiv:1807.06209 [astro-ph.CO]}}. [Erratum: Astron.Astrophys. 652, C4
  (2021)].

\bibitem{Loeb2003}
A.~Loeb and M.~Zaldarriaga, ``{Measuring the small - scale power spectrum of
  cosmic density fluctuations through 21 cm tomography prior to the epoch of
  structure formation}'',
  \href{http://dx.doi.org/10.1103/PhysRevLett.92.211301}{{\em Phys. Rev. Lett.}
  {\bfseries 92} (2004) 211301},
  \href{http://arxiv.org/abs/astro-ph/0312134}{{\ttfamily
  arXiv:astro-ph/0312134}}.

\bibitem{Zaldarriaga2003}
M.~Zaldarriaga, S.~R. Furlanetto, and L.~Hernquist, ``{21 Centimeter
  fluctuations from cosmic gas at high redshifts}'',
  \href{http://dx.doi.org/10.1086/386327}{{\em Astrophys. J.} {\bfseries 608}
  (2004) 622--635}, \href{http://arxiv.org/abs/astro-ph/0311514}{{\ttfamily
  arXiv:astro-ph/0311514}}.

\bibitem{Furlanetto2006}
S.~Furlanetto, S.~P. Oh, and F.~Briggs, ``{Cosmology at Low Frequencies: The 21
  cm Transition and the High-Redshift Universe}'',
  \href{http://dx.doi.org/10.1016/j.physrep.2006.08.002}{{\em Phys. Rept.}
  {\bfseries 433} (2006) 181--301},
  \href{http://arxiv.org/abs/astro-ph/0608032}{{\ttfamily
  arXiv:astro-ph/0608032}}.

\bibitem{Lewis2007}
A.~Lewis and A.~Challinor, ``{The 21cm angular-power spectrum from the dark
  ages}'', \href{http://dx.doi.org/10.1103/PhysRevD.76.083005}{{\em Phys. Rev.
  D} {\bfseries 76} (2007) 083005},
  \href{http://arxiv.org/abs/astro-ph/0702600}{{\ttfamily
  arXiv:astro-ph/0702600}}.

\bibitem{Pritchard2008}
J.~R. Pritchard and A.~Loeb, ``{Evolution of the 21 cm signal throughout cosmic
  history}'', \href{http://dx.doi.org/10.1103/PhysRevD.78.103511}{{\em Phys.
  Rev. D} {\bfseries 78} (2008) 103511},
  \href{http://arxiv.org/abs/0802.2102}{{\ttfamily arXiv:0802.2102
  [astro-ph]}}.

\bibitem{Furlanetto2009}
S.~Furlanetto {\em et~al.}, ``{Cosmology from the Highly-Redshifted 21 cm
  Line}'', \href{http://arxiv.org/abs/0902.3259}{{\ttfamily arXiv:0902.3259
  [astro-ph.CO]}}.

\bibitem{Pritchard2010}
J.~R. Pritchard and A.~Loeb, ``{Constraining the unexplored period between the
  dark ages and reionization with observations of the global 21 cm signal}'',
  \href{http://dx.doi.org/10.1103/PhysRevD.82.023006}{{\em Phys. Rev. D}
  {\bfseries 82} (2010) 023006},
  \href{http://arxiv.org/abs/1005.4057}{{\ttfamily arXiv:1005.4057
  [astro-ph.CO]}}.

\bibitem{Pritchard2011}
J.~R. Pritchard and A.~Loeb, ``{21-cm cosmology}'',
  \href{http://dx.doi.org/10.1088/0034-4885/75/8/086901}{{\em Rept. Prog.
  Phys.} {\bfseries 75} (2012) 086901},
  \href{http://arxiv.org/abs/1109.6012}{{\ttfamily arXiv:1109.6012
  [astro-ph.CO]}}.

\bibitem{Furlanetto2019b}
S.~R. Furlanetto, ``{Physical Cosmology From the 21-cm Line}'',
  \href{http://arxiv.org/abs/1909.12430}{{\ttfamily arXiv:1909.12430
  [astro-ph.CO]}}.

\bibitem{Bernal2022}
J.~L. Bernal and E.~D. Kovetz, ``{Line-intensity mapping: theory review with a
  focus on star-formation lines}'',
  \href{http://dx.doi.org/10.1007/s00159-022-00143-0}{{\em Astron. Astrophys.
  Rev.} {\bfseries 30} no.~1, (2022) 5},
  \href{http://arxiv.org/abs/2206.15377}{{\ttfamily arXiv:2206.15377
  [astro-ph.CO]}}.

\bibitem{Madau1996}
P.~Madau, A.~Meiksin, and M.~J. Rees, ``{21-CM tomography of the intergalactic
  medium at high redshift}'', \href{http://dx.doi.org/10.1086/303549}{{\em
  Astrophys. J.} {\bfseries 475} (1997) 429},
  \href{http://arxiv.org/abs/astro-ph/9608010}{{\ttfamily
  arXiv:astro-ph/9608010}}.

\bibitem{Silk2020}
J.~Silk, ``{The limits of cosmology: role of the Moon}'',
  \href{http://dx.doi.org/10.1098/rsta.2019.0561}{{\em Phil. Trans. A. Math.
  Phys. Eng. Sci.} {\bfseries 379} (2021) 20190561},
  \href{http://arxiv.org/abs/2011.04671}{{\ttfamily arXiv:2011.04671
  [astro-ph.CO]}}.

\bibitem{Munoz2015}
J.~B. Mu\~noz, Y.~Ali-Ha\"\i{}moud, and M.~Kamionkowski, ``{Primordial
  non-gaussianity from the bispectrum of 21-cm fluctuations in the dark
  ages}'', \href{http://dx.doi.org/10.1103/PhysRevD.92.083508}{{\em Phys. Rev.
  D} {\bfseries 92} no.~8, (2015) 083508},
  \href{http://arxiv.org/abs/1506.04152}{{\ttfamily arXiv:1506.04152
  [astro-ph.CO]}}.

\bibitem{Amon2022}
A.~Amon and G.~Efstathiou, ``{A non-linear solution to the $S_8$ tension?}'',
  \href{http://arxiv.org/abs/2206.11794}{{\ttfamily arXiv:2206.11794
  [astro-ph.CO]}}.

\bibitem{Preston2023}
C.~Preston, A.~Amon, and G.~Efstathiou, ``{A non-linear solution to the $S_8$
  tension II: Analysis of DES Year 3 cosmic shear}'',
  \href{http://arxiv.org/abs/2305.09827}{{\ttfamily arXiv:2305.09827
  [astro-ph.CO]}}.

\bibitem{Schive2014}
H.-Y. Schive, T.~Chiueh, and T.~Broadhurst, ``{Cosmic Structure as the Quantum
  Interference of a Coherent Dark Wave}'',
  \href{http://dx.doi.org/10.1038/nphys2996}{{\em Nature Phys.} {\bfseries 10}
  (2014) 496--499}, \href{http://arxiv.org/abs/1406.6586}{{\ttfamily
  arXiv:1406.6586 [astro-ph.GA]}}.

\bibitem{MarshPop2015}
D.~J.~E. Marsh and A.-R. Pop, ``{Axion dark matter, solitons and the
  cusp\textendash{}core problem}'',
  \href{http://dx.doi.org/10.1093/mnras/stv1050}{{\em Mon. Not. Roy. Astron.
  Soc.} {\bfseries 451} no.~3, (2015) 2479--2492},
  \href{http://arxiv.org/abs/1502.03456}{{\ttfamily arXiv:1502.03456
  [astro-ph.CO]}}.

\bibitem{Mocz2023}
P.~Mocz {\em et~al.}, ``{Cosmological Structure Formation and Soliton Phase
  Transition in Fuzzy Dark Matter with Axion Self-Interactions}'',
  \href{http://arxiv.org/abs/2301.10266}{{\ttfamily arXiv:2301.10266
  [astro-ph.CO]}}.

\bibitem{Lague2023}
A.~Lagu\"e, B.~Schwabe, R.~Hlo\v{z}ek, D.~J.~E. Marsh, and K.~K. Rogers,
  ``{Cosmological simulations of mixed ultralight dark matter}'',
  \href{http://arxiv.org/abs/2310.20000}{{\ttfamily arXiv:2310.20000
  [astro-ph.CO]}}.

\bibitem{Blas2011}
D.~Blas, J.~Lesgourgues, and T.~Tram, ``{The Cosmic Linear Anisotropy Solving
  System (CLASS) II: Approximation schemes}'',
  \href{http://dx.doi.org/10.1088/1475-7516/2011/07/034}{{\em JCAP} {\bfseries
  07} (2011) 034}, \href{http://arxiv.org/abs/1104.2933}{{\ttfamily
  arXiv:1104.2933 [astro-ph.CO]}}.

\bibitem{Scelfo2018}
G.~Scelfo, N.~Bellomo, A.~Raccanelli, S.~Matarrese, and L.~Verde,
  ``{GW$\times$LSS: chasing the progenitors of merging binary black holes}'',
  \href{http://dx.doi.org/10.1088/1475-7516/2018/09/039}{{\em JCAP} {\bfseries
  09} (2018) 039}, \href{http://arxiv.org/abs/1809.03528}{{\ttfamily
  arXiv:1809.03528 [astro-ph.CO]}}.

\bibitem{Bellomo2020}
N.~Bellomo, J.~L. Bernal, G.~Scelfo, A.~Raccanelli, and L.~Verde, ``{Beware of
  commonly used approximations. Part I. Errors in forecasts}'',
  \href{http://dx.doi.org/10.1088/1475-7516/2020/10/016}{{\em JCAP} {\bfseries
  10} (2020) 016}, \href{http://arxiv.org/abs/2005.10384}{{\ttfamily
  arXiv:2005.10384 [astro-ph.CO]}}.

\bibitem{Scelfo2021}
G.~Scelfo, M.~Spinelli, A.~Raccanelli, L.~Boco, A.~Lapi, and M.~Viel,
  ``{Gravitational waves \texttimes{} HI intensity mapping: cosmological and
  astrophysical applications}'',
  \href{http://dx.doi.org/10.1088/1475-7516/2022/01/004}{{\em JCAP} {\bfseries
  01} no.~01, (2022) 004}, \href{http://arxiv.org/abs/2106.09786}{{\ttfamily
  arXiv:2106.09786 [astro-ph.CO]}}.

\bibitem{Hall2013}
A.~Hall, C.~Bonvin, and A.~Challinor, ``{Testing General Relativity with 21-cm
  intensity mapping}'',
  \href{http://dx.doi.org/10.1103/PhysRevD.87.064026}{{\em Phys. Rev. D}
  {\bfseries 87} no.~6, (2013) 064026},
  \href{http://arxiv.org/abs/1212.0728}{{\ttfamily arXiv:1212.0728
  [astro-ph.CO]}}.

\bibitem{Shiraishi2016}
M.~Shiraishi, J.~B. Mu\~noz, M.~Kamionkowski, and A.~Raccanelli, ``{Violation
  of statistical isotropy and homogeneity in the 21-cm power spectrum}'',
  \href{http://dx.doi.org/10.1103/PhysRevD.93.103506}{{\em Phys. Rev. D}
  {\bfseries 93} no.~10, (2016) 103506},
  \href{http://arxiv.org/abs/1603.01206}{{\ttfamily arXiv:1603.01206
  [astro-ph.CO]}}.

\bibitem{Furlanetto2019a}
S.~Furlanetto {\em et~al.}, ``{Astro 2020 Science White Paper: Fundamental
  Cosmology in the Dark Ages with 21-cm Line Fluctuations}'',
  \href{http://arxiv.org/abs/1903.06212}{{\ttfamily arXiv:1903.06212
  [astro-ph.CO]}}.

\end{thebibliography}\endgroup
\bibliographystyle{utcaps}

\end{document}